\documentclass{aa}  

\usepackage{graphicx}
\usepackage{xcolor}
\usepackage{natbib}
\usepackage{ulem}
\usepackage{txfonts}
\usepackage{floatpag}
\usepackage{hyperref}
\usepackage{gensymb} 
\newcommand{\chieff}{\ensuremath{\chi_\mathrm{eff}}\xspace}
\newcommand{\mchirp}{\ensuremath{\mathrm{M}_\mathrm{chirp}}\xspace}
\newcommand{\rdunits}{\ensuremath{\mathrm{Gpc}^{-3}\,\mathrm{yr}^{-1}}\xspace}

\newcommand{\cosmic}{{\texttt{COSMIC}\xspace{}}}
\newcommand{\mesa}{{\texttt{MESA}\xspace{}}}
\newcommand{\posydon}{{\texttt{POSYDON}\xspace{}}}
\newcommand{\compas}{{\texttt{COMPAS}\xspace{}}}

\definecolor{chmagenta}{rgb}{0.54, 0.17, 0.88}

\definecolor{midnightblue}{rgb}{0.1, 0.1, 0.44}

\usepackage{soul}

\begin{document} 

   \title{The impact of mass-transfer physics on the observable properties of field binary black hole populations}
   \titlerunning{The impact of mass-transfer physics on the observable properties of field BBH populations}
   
   \author{Simone\,S.\,Bavera\inst{1}\fnmsep\thanks{E-mail: simone.bavera@unige.ch},
          Tassos\,Fragos\inst{1},
          Michael\,Zevin\inst{2,3}\fnmsep\thanks{NASA Hubble Fellow},
          Christopher\,P.\,L.\,Berry\inst{2,4},
          Pablo\,Marchant\inst{5},
          Jeff\,J.\,Andrews\inst{2},
          Scott\,Coughlin\inst{2},
          Aaron\,Dotter\inst{2},
          Konstantinos\,Kovlakas\inst{6,7},
          Devina\,Misra\inst{1},
          Juan\,G.\,Serra-Perez\inst{2},
          Ying\,Qin\inst{2},
          Kyle\,A.\,Rocha\inst{2},
          Jaime\,Román-Garza\inst{1},
          Nam\,H.\,Tran\inst{8} \and
          Emmanouil\,Zapartas\inst{1}
          }
   \authorrunning{S. S. Bavera et al.}
   
   \institute{Geneva Observatory, University of Geneva, Chemin des Maillettes 51, 1290 Versoix, Switzerland
              \and
              Center for Interdisciplinary Exploration and Research in Astrophysics (CIERA) and Department of Physics and Astronomy, Northwestern University, 1800 Sherman Avenue, Evanston, IL 60201, USA
              \and
              Enrico Fermi Institute and Kavli Institute for Cosmological Physics, The University of Chicago, 5640 South Ellis Avenue, Chicago, Illinois 60637, USA
              \and
              SUPA, School of Physics and Astronomy, University of Glasgow, Glasgow G12 8QQ, UK
              \and
              Institute of Astrophysics, KU Leuven, Celestijnenlaan 200D, 3001, Leuven, Belgium
              \and
              Physics Department, University of Crete, GR 71003, Heraklion, Greece
              \and
              Institute of Astrophysics, Foundation for Research and Technology-Hellas, GR 71110 Heraklion, Greece
              \and
              DARK, Niels Bohr Institute, University of Copenhagen, Jagtvej 128, 2200 Copenhagen, Denmark
             }

   \date{Accepted January 25, 2021}

  \abstract
   {We study the impact of mass-transfer physics on the observable properties of binary black hole populations that formed through isolated binary evolution. We used the \posydon{} framework to combine detailed \mesa{} binary simulations with the \cosmic{} population synthesis tool to obtain an accurate estimate of merging binary black hole observables with a specific focus on the spins of the black holes. We investigate the impact of mass-accretion efficiency onto compact objects and common-envelope efficiency on the observed distributions of the effective inspiral spin parameter \chieff, chirp mass \mchirp, and binary mass ratio $q$. We find that low common envelope efficiency translates to tighter orbits following the common envelope and therefore more tidally spun up second-born black holes. However, these systems have short merger timescales and are only marginally detectable by current gravitational-wave detectors as they form and merge at high redshifts ($z \sim 2$), outside current detector horizons. Assuming Eddington-limited accretion efficiency and that the first-born black hole is formed with a negligible spin, we find that all non-zero $\chi_\mathrm{eff}$ systems in the detectable population can come only from the common envelope channel as the stable mass-transfer channel cannot shrink the orbits enough for efficient tidal spin-up to take place. We find that the local rate density ($z\simeq 0.01$) for the common envelope channel is in the range of $\sim 17$--$113~\mathrm{Gpc^{-3}\,yr^{-1}}$, considering a range of $\alpha_\mathrm{CE} \in [0.2,5.0]$, while for the stable mass transfer channel the rate density is $\sim 25~\mathrm{Gpc^{-3}\,yr^{-1}}$. The latter drops by two orders of magnitude if the mass accretion onto the black hole is not Eddington limited because conservative mass transfer does not shrink the orbit as efficiently as non-conservative mass transfer does. Finally, using GWTC-2 events, we constrained the lower bound of branching fraction from other formation channels in the detected population to be $\sim 0.2$. Assuming all remaining events to be formed through either stable mass transfer or common envelope channels, we find moderate to strong evidence in favour of models with inefficient common envelopes.
   }


   \maketitle
%

\section{Introduction}\label{sec:introduction}

Stars in binary systems are common in the Universe \citep{2012Sci...337..444S}, but the details of their evolution are uncertain. For massive binaries, it is difficult to observationally constrain the details of physical processes, such as mass-transfer (MT), as the lifetimes of these interacting binary phases are short and hence it is unlikely to observe many of them directly. However, with gravitational-wave (GW) observations, one can search for the imprints of these processes on the properties of their stellar remnant populations, such as the binary black hole (BBH) population. 

The LIGO Scientific and Virgo Collaboration (LVC) has recently released the new GW catalogue GWTC-2 \citep{abbott2020gwtc2}, which includes 37 new potential BBH detections\footnote{Here we consider GW190814 as a possible BBH.} from the first half of the third observing run (O3a). In total, GWTC-2 contains 47 BBHs detections \citep{2019PhRvX...9c1040A,abbott2020gwtc2}, and the intrinsic rate density of BBH mergers is currently estimated to be $23.9^{+14.9}_{-8.6}$~\rdunits  \citep{theligoscientificcollaboration2020population}.
Each GW detection can constrain some combination of the astrophysical BH parameters: spin and mass. Convenient well-measured quantities are the chirp mass
\begin{equation}
    \mchirp = \frac{(m_1 m_2)^{3/5}}{(m_1+m_2)^{1/5}} \, ,
\end{equation}
where $m_1$ and $m_2$ ($m_1 \geq m_2$) are the BHs masses, the binary mass ratio $q=m_2/m_1 \leq 1$, and the effective inspiral spin parameter
\begin{equation}
    \chieff = \frac{m_1 \vec{a}_1 + m_2 \vec{a}_2}{m_1 + m_2} \cdot \vec{\hat{L}} \, ,
\end{equation}
where $\vec{\hat{L}}$ is the orbital angular momentum (AM) unit vector and $\vec{a}_1$ and $\vec{a}_2$ are the BH dimensionless spin vectors. The dimensionless spin vectors are defined as
\begin{equation}
    \vec{a_i} = \frac{c \vec{J}_i}{G m_i^2} \, , \, \, \, i \in \{1,2\} \, ,
\end{equation}
where $c$ is the speed of light, $G$ is the gravitational constant and $\vec{J}_i$ is the spin AM vector of the BH. There is a degeneracy between \chieff and $q$ which limits the accuracy to which each quantity can be measured independently \citep{1995PhRvD..52..848P,2013ApJ...766L..14H}. Nevertheless, the combination of the three observables provide a robust constraint on the properties of a BBH.

Multiple formation channels have been proposed to explain the origin of merging BBHs. They can be divided into two broad categories: (i) isolated binary evolution and (ii) dynamical assembly.

The former occurs during isolated stellar evolution in the field under some specific binary evolution interactions. Interacting binaries that after the formation of the first BH go through (A) stable mass transfer \citep[SMT; e.g.][]{2017MNRAS.471.4256V,2017MNRAS.468.5020I,2019MNRAS.490.3740N} or (B) unstable mass transfer leading to a common envelope (CE) phase \citep[e.g.][]{1976ApJ...207..574S,1976IAUS...73...35V,1993MNRAS.260..675T,2007PhR...442...75K,2014LRR....17....3P,2016Natur.534..512B} have been shown to form merging BBHs. Another possibility is the formation of BBHs from massive stars with low metallicities and orbital period less than $\sim$4 days, which due to their tidal interaction, can maintain an almost critical rotation and are going to evolve (C) chemically homogeneously \citep[e.g.][]{2009A&A...497..243D,2016MNRAS.458.2634M,2016A&A...588A..50M,2020MNRAS.tmp.3003D}.  

The second category of formation channels category occurs in dense stellar environments where stars and binaries can dynamically interact with each other and assemble new binary systems with more massive BHs and tighter orbits, that may eventually merge within the Hubble time. This formation path is present in (D) globular, open, and nuclear stellar clusters \citep[e.g.][]{1993Natur.364..421K,1993Natur.364..423S,2000ApJ...528L..17P,2009ApJ...692..917M,2010MNRAS.402..371B,2015PhRvL.115e1101R,2016ApJ...831..187A,2018MNRAS.477.4423A,2018PhRvL.121p1103F,2019PhRvD.100d3027R} and (E) active galactic nuclei disks \citep[e.g.][]{2017ApJ...835..165B,2017MNRAS.464..946S,2018ApJ...866...66M,2020ApJ...899...26T}. 
Finally (F) triple or higher-order stellar systems can also lead to the formation of BBHs \citep[e.g.][]{2017ApJ...836...39S,2018ApJ...863....7R,2020PhRvD.101j4053G,2020A&A...640A..16T}.
Within their uncertainties, almost all of these formation channels have been shown to have rate estimates consistent or marginally consistent with the empirical LVC rates.

In this study we consider the formation of BBHs in isolated binary evolution (A and B) though the SMT and CE phase. In these formation channels, two massive stars are born in a relatively wide binary (orbital separations of order $\sim 1000\,\rm \mathrm{R}_{\odot}$), where binary interactions happen after the more massive star leaves the main sequence (MS). At this stage, the star expands to become a red supergiant, and inflates its hydrogen-rich envelope beyond its Roche lobe, leading to the first MT episode. The MT stops when the entire stellar envelope is lost, leaving behind a naked helium (He)-star which eventually collapses to form a BH. When the companion reaches the end of its MS, the process repeats itself for the companion star. This MT phase can be either stable or unstable, with the latter leading to the formation of a CE of gas engulfing the binary. If the stripping of the secondary's envelope is successful, we are left with either a tight BH--He-star system in the case of CE, or with a somewhat wider system in the case of SMT. Eventually the secondary star also collapses to form the second-born BH, and due to energy and AM loss from GW emission \citep{1964PhRv..136.1224P}, the BBH system coalesces to form a single, more massive BH.

In the SMT and CE formation channels, the spin of the first-born BH is determined by the AM transport efficiency during the evolution of the progenitor star. Measurements of neutron star and white dwarf spins \citep{2005ApJ...626..350H,2008A&A...481L..87S} and asteroseismology studies \citep{2014ApJ...796...17F,2014ApJ...788...93C} suggest that this mechanism must be efficient \citep{1999A&A...349..189S,2002A&A...381..923S,2019ApJ...881L...1F}. Thus, upon expansion, the initial AM of the star is mostly transported to the outer layers which are subsequently lost due to MT and wind mass loss. This leads to the formation of slowly spinning BHs ($a_1 \simeq 0$), as initially suggested in the context of BH low-mass X-ray binary formation by \citet{2015ApJ...800...17F} and subsequently quantitatively shown in \citet{2018A&A...616A..28Q}, \citet{2019ApJ...881L...1F} and \citet{2020A&A...636A.104B}. In the case of the SMT channel, during the second MT episode, the first born BH may accrete material and spin up \citep{1974ApJ...191..507T}, depending on the accretion efficiency rate. On the other hand, the spin of the second-born BH is determined by the net effect of the stellar wind and the tidal interaction of the BH--He-star binary system. Because of the efficiency of the AM transport, the He-star emerges from the second MT event with a negligible spin. If the orbital separation is small enough and stellar winds do not widen the system significantly, the He-star can be spun up by tides. These conditions are met at low metallicities for BBHs formed through the CE formation channel \citep{2020A&A...635A..97B}. In contrast, in the case of SMT, the orbits shrink less efficiently leading to less tidally spun up second-born BHs compared to the CE channel. 

All formation channels can be investigated through population synthesis studies which adopt stellar and binary models to rapidly evolve millions of binary stars. This approach gives us insights on the overall population observables given a set of physical assumptions. To explore a wide landscape of parameter values and generate many realisations of the studied population, we need to efficiently evolve millions of binaries. This can be achieved through parametric population synthesis codes which employ fits of single stellar evolution with analytical models to simulate the binary interactions at the expense of coarser approximations when modelling these interactions. Despite this limitation, rapid population synthesis still allows for investigation into how the observable distributions of a population change within the astrophysical model uncertainties. \citet{2020A&A...635A..97B} recently showed how, given a specific theoretical framework, one can adopt detailed stellar and binary simulations in a population synthesis study to obtain new observable estimates such as the BH spin distributions. In this paper we study how these model predictions are affected by the uncertainties of MT physics such as MT stability and efficiency, CE efficiency and initial orbital distributions.

In Sec.~\ref{sec:methods} we present the framework we use to generate the population of BBHs and how we convolve the synthetic BBH population with the redshift- and metallicity-dependent star formation rate, as well as incorporate GW detector selection effects. We also summarise the key differences between this work and \citet{2020A&A...635A..97B}. The BBH observable distributions for different MT and CE efficiencies are presented in Sec.~\ref{sec:results}, where we also show how the \chieff, \mchirp and $q$ distributions change for these different physical assumptions and determinate thorough model selection which CE efficiency is supported by GWTC-2 BBHs events. The impact of MT stability and initial orbital distributions on the uncertainties of our models is discussed in Sec.~\ref{sec:modeluncert.}. We conclude by summarising our findings in Sec.~\ref{sec:conclusions}.

\section{Methods}\label{sec:methods}

We generate our populations of BBHs by modelling isolated binary evolution with the \posydon{} code.\footnote{See Fragos et al. (2021), to be submitted by the POSYDON collaboration, \href{http://www.posydon.org}{www.posydon.org}.} 
\posydon, among many other functionalities, can run and combine detailed stellar and binary evolution simulations performed with the \mesa{} code \citep{2011ApJS..192....3P,2013ApJS..208....4P, 2015ApJS..220...15P,2018ApJS..234...34P,2019ApJS..243...10P} to existing parametric binary population synthesis codes. This integration lets us target particular evolutionary phases with more detailed modelling. Similar hybrid approaches have been used in previous population synthesis studies \citep[e.g.][]{2012JPhCS.341a2008N,2014MNRAS.445.1912C,2015ApJ...802L...5F,2015ApJ...802..131S,2019ApJ...886..118S}. In this work, we use \cosmic{} \citep{2020ApJ...898...71B} to model the evolution of binaries starting from zero age MS (ZAMS) until the formation of the BH--He-star system. We then use \mesa{} to model in detail the subsequent evolution until the formation of the BBH which is the evolutionary phase that determines the spin of the second-born BH \citep{2018A&A...616A..28Q,2020A&A...635A..97B}. 

\subsection{Binary black hole population}\label{sec:BBHpop}

We create synthetic BBH populations similar to \citet{2020A&A...635A..97B}, but with some key differences.

We assume similar initial binary properties with the exception of the initial orbital periods which here are drawn from an extended \citet{2012Sci...337..444S} log-power law with coefficient $\pi = - 0.55$ in the range $p \in [10^{0.15},10^{5.5}]$ days and extrapolated down to $p = 0.4$ days assuming a log-flat distribution (as the power law is not defined for $p < 10^{0.15}$ day). This extension includes the portion of the parameter space leading to chemical homogeneous evolution \citep{2020MNRAS.tmp.3003D}. All initial binary property assumptions are explained in Appendix~\ref{app:initial_properties} and discussed in Sec~\ref{sec:initalprop}.

To determine the MT stability, we adopt critical mass ratios $q_\mathrm{crit}$ as in \citet{2019MNRAS.490.3740N} with one exception. For stars in the giant branch (GB) and asymptotic giant branch (AGB) we use the same $q_\mathrm{crit}$  fits as in \citet{2019MNRAS.490.3740N} but do not adopt \citet{1997A&A...327..620S} radial response to adiabatic mass loss for evolved stars beyond the Hertzsprung gap (HG) because they are not currently available in \texttt{COSMIC}. For our reference models, the stable mass-accretion efficiency onto degenerate objects is Eddington-limited. This leads to a highly non-conservative mass-transfer phase where the first-born BH accretes a negligible amount of matter and cannot spin up due to accretion. Unstable MT is parameterised with the standard $\alpha_\mathrm{CE}$--$\lambda$ formalism \citep[see e.g.][for a review]{2013A&ARv..21...59I}. In contrast to \citet{2020A&A...635A..97B}, we adopt $\lambda$ fits as in \citet{2014A&A...563A..83C} while we explore different $\alpha_\mathrm{CE}$ efficiencies: $\alpha_\mathrm{CE} \in [0.2,0.35,0.5,0.75,1.0,2.0,5.0]$. Since, approximately, $\alpha_\mathrm{CE}$ scales linearly with the orbital separation post CE, see Eq.~\eqref{eq:CE}, low CE efficiencies lead to tighter orbital separations post CE. Therefore, we expect that more BH--He-star systems will undergo tidal spin up at lower $\alpha_\mathrm{CE}$. We describe the details of our \cosmic{} model, MT stability and CE in Appendix~\ref{app:COSMIC}. In Sec.~\ref{sec:MTstability} and ~\ref{sec:qcrit} we discuss how our BBH population distributions and rates are affected by these assumptions.

The late-end phase of the binary evolution of BBHs formed through CE and SMT channels are BH--He-star systems. We update our \mesa{} models \citep{2018A&A...616A..28Q,2020A&A...635A..97B} to match the stellar model assumptions of \citet{2020MNRAS.tmp.3003D}. In contrast to \citet{2020A&A...635A..97B}, we relax the He-star models to zero age helium MS (ZAHeMS) before initiating the binary evolution. This ensures that the He-star model is in thermal and hydrostatic equilibrium when the binary interactions begin. In order to verify that the He-star will not overfill the L2 Roche volume throughout the binary evolution, we include the prescription of \cite{2020A&A...642A.174M}. The ingredients of our \mesa{} model are explained in Appendix~\ref{app:MESA}.

Once the He-star reaches carbon depletion, the \mesa{} simulations are stopped. We then collapsed the profile of the He-star according to the procedure used in \citet{2020A&A...635A..97B} which accounts for disk formation. Here we adopted a different treatment of neutrino mass loss where we assume that the innermost $3~\mathrm{M}_\odot$ forms a proto-neutron star which collapses to form a BH of $2.5~\mathrm{M}_\odot$ while $0.5~\mathrm{M}_\odot$ are converted into neutrinos and escape the system carrying away AM \citep[cf.][]{2020ApJ...899L...1Z}. The complete procedure used to collapse the He-star profiles is explained in detail in Appendix~\ref{app:corecollapse}.

\begin{figure}[!htbp]
\centering
\includegraphics[width=0.98\columnwidth]{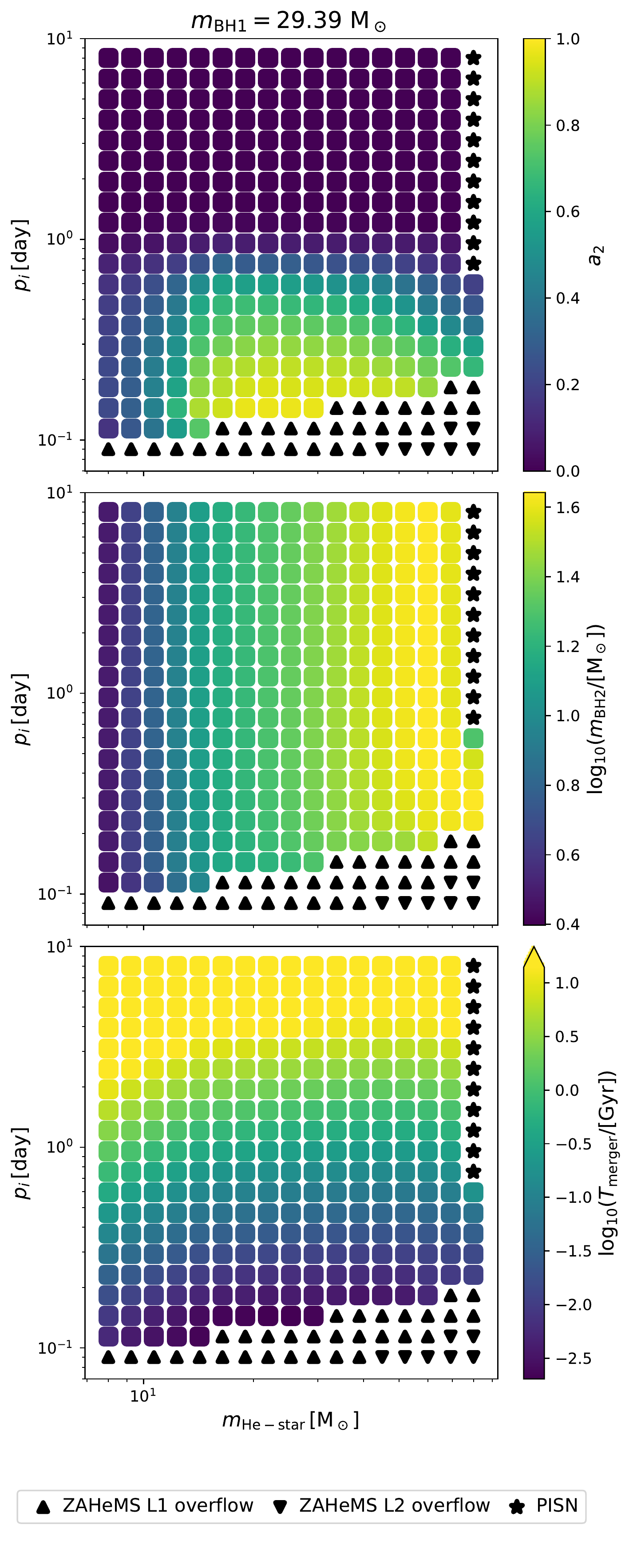}
 \caption{Example of a two-dimensional slice of the four-dimensional grid showing their initial BH--He-star orbital period $p_i$ (in days) and initial He-star mass $m_\mathrm{He-star}$ (in M$_\odot$) for $\log_{10}(Z) \simeq -3$ and $m_\mathrm{BH1} = 29.39 \, \mathrm{M}_\odot$. The final mass $m_\mathrm{BH2}$ and spin $a_2$ of the second-born BH, as well as merger timescale $T_\mathrm{merger}$, are coloured according to the legend of each panel. All successful \mesa{} simulation stopped at carbon depletion (square markers) while other termination flags are shown in the bottom legend. The merger timescale colour bar is capped at $14~\mathrm{Gyr}$.}
 \label{fig:grids-small} 
\end{figure}
   
We use our detailed binary stellar models to cover the four-dimensional parameter space of initial metallicity $Z$, BH mass $M_\mathrm{BH}$, He-star mass $M_\mathrm{He-star}$ and orbital period $p$. We run grids for $30$ different metallicities ranging from $\log_{10}(Z) = -4.0$ to $\log_{10}(1.5 Z_\odot) \simeq -1.593$ in steps of $\log_{10}(Z) \simeq 0.083$ where we adopt the solar reference $Z_\odot = 0.017$ \citep{1996ASPC...99..117G}. For each metallicity we run $11$ BH masses in the log-range $[2.5, 54.4] \, \mathrm{M}_\odot$, $17$ He-star masses in the log-range $[8, 80] \, \mathrm{M}_\odot$ and $20$ initial binary periods in the log-range $[0.09, 8] \, \mathrm{days}$. In total, we calculated roughly $110,000$ new binary evolution sequences. These grids were used to determine the final outcomes and final parameters of the late-end evolution stage of the binary systems through linear interpolation for each metallicity, independently. The features of these grids and the interpolation accuracy are discussed in Appendix~\ref{app:MESA-grids}.

In Fig.~\ref{fig:grids-small} we show an example of a two-dimensional slice from our four-dimensional grid. The parameter space is sliced at $Z=0.001$ and $M_\mathrm{BH}= 29.4 \, \mathrm{M}_\odot$. We show the final second-born BH mass and spin, as well as the BBH merger timescale as a function of initial orbital period and He-star mass at ZAHeMS. We see that the binary interactions determining the spin of the second-born BH create a gradual transition between tidally locked systems and non-spinning systems. The complex interactions of stellar winds, tides, internal differential rotation and (in some cases) mass transfer are important in determining the spin of the second-born BH, and therefore require a detailed treatment which traditional rapid population codes cannot offer. 

Once a BBH system is formed, GW inspiral leads to the system's eventual merger. We calculate the merger timescale $T_\mathrm{merger}$ as in \citet{1964PhRv..136.1224P} accounting for eccentricity. In our models, this timescale is anti-correlated with the observable $\chi_\mathrm{eff}$. This is caused by tides, as they are the only mechanism able to spin up the progenitor of the second-born BH. Their efficiency is highly dependent on the orbital separation, see Eq.~\eqref{eq:tides}. For tidally locked systems one recovers $T_\mathrm{merger} \propto a_2^{-8/3} \propto \chi_\mathrm{eff}^{-8/3}$ while for the other systems $T_\mathrm{merger} \propto a_2^{-8/17} \propto \chi_\mathrm{eff}^{-8/17}$ \citep{2020A&A...635A..97B}. We therefore expect systems with $\chieff > 0$ to have, on average, shorter merger timescales compared to systems with $\chieff = 0$. This anti-correlation is the key to understanding the translation from the underlying BBH population (what one would observe with an infinitely sensitive detector) to the observed population. Current GW detectors are probing small redshifts ($z \lesssim 1$) and cannot explore the peak of the Universe's SFR at $z \simeq 2$ where most of the highly spinning BBHs form and merge  \citep[as they are preferentially formed in low metallicity environments, e.g.][]{2017NatCo...814906S}.

\subsection{Rate estimates}\label{sec:weights}

To compute the expected rate of detectable GW events, we need
to convolve the redshift- and metallicity-dependent star-formation rate (SFR) with the selection effects of the detector array. 
To do this, we follow the approach shown in Appendix~B of \citet{2020A&A...635A..97B}. We assume a flat $\Lambda$CDM cosmology with $H_0 = 67.7~\mathrm{km\,s^{-1}\,Mpc^{-1}}$ and $\Omega_m=0.307$ \citep{2016A&A...594A..13P}, a cosmic SFR history as in \citet{2017ApJ...840...39M} and metallicities following a truncated log-normal distribution with standard deviation $0.5~\mathrm{dex}$ around the empirical mean metallicity function derived by \citet{2017ApJ...840...39M}.
The log-normal distribution is truncated at the highest metallicity bin edge and the distribution is accordingly renormalised to ensure that $\int_{- \infty}^{Z_\mathrm{max}} N(\log_{10}(Z) \, | \, \mu(z),\sigma) \, \mathrm{d} \log_{10} Z = 1$, where $Z_\mathrm{max}$ is our highest metallicity edge bin. Portions of the distribution extending beyond the lower limit edge are included in the edge bin when integrating over metallicity.
The population synthesis predictions are performed in finite time bins of $\Delta t_i = 100~\mathrm{Myr}$ and log-metallicity bins $\Delta Z_j$. 
The detection rate of BBH mergers for a given detector network is calculated from the Monte Carlo simulations, cf. Eq.~(13) in \citet{2020A&A...635A..97B},
\begin{equation}
    R_\mathrm{det} = \sum_{\Delta t_i}  \sum_{\Delta Z_j} \sum_{k} f_\mathrm{corr} \frac{\mathrm{fSFR}(z_{\mathrm{f},i})}{M_{\mathrm{sim}, \, \Delta Z_j}} 4 \pi c \, D^2_\mathrm{c}(z_{\mathrm{m},i,k}) \, p_{\mathrm{det},i,k} \, \Delta t_i \, \, \, \mathrm{yr^{-1}},
    \label{eq:Rdet}
\end{equation}
where the argument of the summation is the cosmological weight contribution of the $k$-th binary born at redshift $z_{\mathrm{f},i}$ with BH mass $m_{1,k}$ and $m_{2,k}$, spin $\vec{a}_{1,k}$ and $\vec{a}_{2,k}$ and merging at redshift $z_{\mathrm{m},i,k}$. Furthermore, $M_{\mathrm{sim},\Delta Z_j}$ is the simulated mass per log-metallicity bin $\Delta Z_j$ and $f_\mathrm{corr}$ the normalisation constant which converts the simulated mass to the total stellar population \citep[see Appendix~A in][]{2020A&A...635A..97B}. Here, $D_\mathrm{c}(z)$ is the comoving distance to the source, $\mathrm{fSFR}(z)$ is the SFR per log-metallicity range $\Delta Z_j$ and $p_\mathrm{det,i,k} \equiv p_\mathrm{det}(z_{\mathrm{m},i,k},m_{1,k},m_{2,k},\vec{a}_{1,k},\vec{a}_{2,k}) $ accounts for the selection effects of the detector array.

In contrast to \citet{2020A&A...635A..97B}, we calculate the sensitivity of a GW detector to a source accounting for its network configuration as well as include the selection effects on the BH spins. We assume a 3-detector network configuration composed of LIGO--Hanford, LIGO--Livingston, and Virgo with simulated O3 sensitivity (mid high/late low from \citealt{2018LRR....21....3A}). 
Detector response functions are calculated using the \texttt{PyCBC} package \citep{alex_nitz_2020_3904502}. 
For each compact binary merger, we calculate the signal-to-noise ratio (S$/$R) as 
\begin{equation}
\rho^2= 4 \Re \int_0^\infty \frac{\tilde{h}^{\star}(f)\tilde{h}(f)}{S_n(f)} \mathrm{d}f
\end{equation}
for each detector in the network, where $S_n(f)$ is the one-sided power spectral density of the noise, and $\tilde{h}(f)$ is the GW strain, determined using the \texttt{IMRPhenomPv2} waveform approximant \citep{PhysRevLett.113.151101,2016PhRvD..93d4007K}. 
The network S$/$R is the quadrature sum of the S$/$Rs in all the three detectors. 
Assuming a network detection threshold of $\rho_{\rm det} = 12$, we Monte Carlo sample the sky location, inclination, and phase $N$ times for each system and calculate $\rho_{\rm net}$. 
The detection probably $p_{\mathrm{det},i,k}$ is thus determined as 
\begin{equation}
    p_{\mathrm{det},i,k} =  \frac{1}{N} \sum^{N}_{l=1} \mathcal{H}(\rho_{\mathrm{net},\,l,i,k} - \rho_{\rm det}) ,
\end{equation}
where each $l$ represents a random draw of extrinsic parameters and $\mathcal{H}$ is the Heaviside step function; 
we perform $N=1000$ realisations of extrinsic parameters for each system. 

The total BBH merger rate density $R_\mathrm{BBH}(z)$ is the number of BBHs per comoving volume per year as a function of redshift. This quantity can be calculated knowing the contribution of each binary $k$ placed at the centre of each formation time bin $\Delta t_i$ in its corresponding metallicity bin $\Delta Z_j$ assuming $p_\mathrm{det,i,k}=1$,
\begin{equation}
    R_\mathrm{BBHs} (z_i) = \sum_{\Delta Z_j} \sum_{k} f_\mathrm{corr} \frac{\mathrm{fSFR}(z_{\mathrm{f},i})}{M_{\mathrm{sim,} \, \Delta Z_j}}
    \frac{4 \pi c \, D^2_\mathrm{c}(z_{\mathrm{m},i,k})}{\Delta V_\mathrm{c}(z_i)} \, \Delta t_i \, \, \, \mathrm{Gpc^{-3} yr^{-1}},
    \label{eq:RBBHs}
\end{equation}
where $\Delta V_\mathrm{c}(z)$ is the comoving volume shell corresponding to the cosmic time bin $\Delta t_i$,
\begin{equation}
    \Delta V_\mathrm{c}(z_i) \equiv \int_{\Delta z_i} \frac{1}{1+z} \frac{\mathrm{d} V_\mathrm{c}}{\mathrm{d}z} \mathrm{d}z = \frac{4\pi c}{H_0} \int_{\Delta z_i} \frac{D_\mathrm{c}^2(z)}{E(z)(1+z)} \mathrm{d}z \, ,
\end{equation}
where $E(z) = \sqrt{\Omega_m(1+z)^3+\Omega_\Lambda}$. Here $\Delta z_i$ is the redshift interval corresponding to the cosmic time bin $\Delta t_i$ centered at $z_i \equiv z_{f,i}$.


   \begin{figure*}
   \centering
      \includegraphics[width=0.9\textwidth]{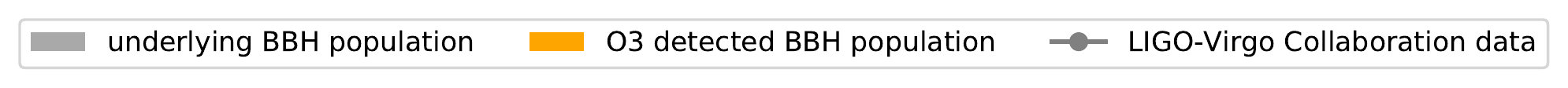} \\
    \begin{minipage}[b]{0.33\textwidth}
    \includegraphics[width=\textwidth]{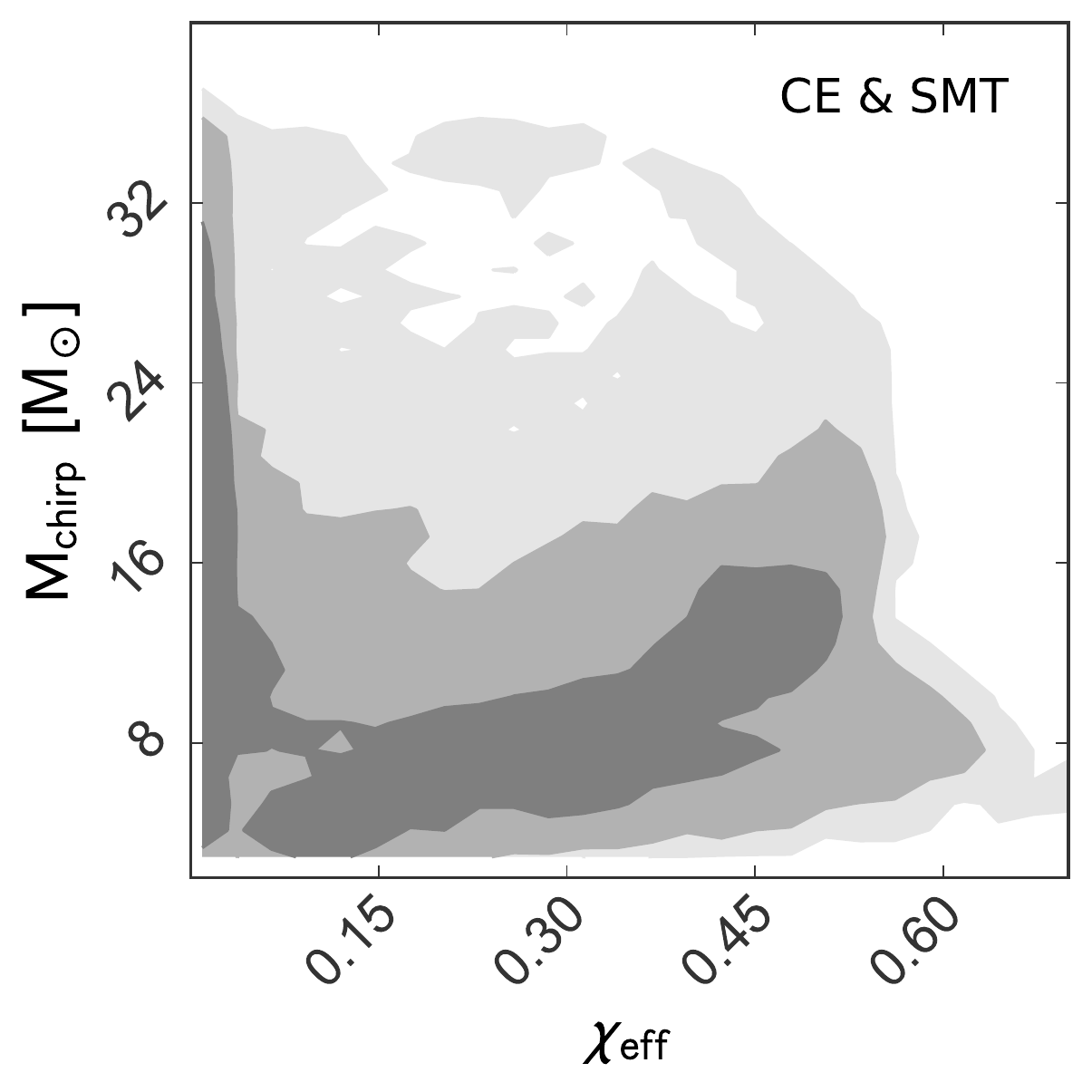}
    \end{minipage}
    \begin{minipage}[b]{0.33\textwidth}
    \includegraphics[width=\textwidth]{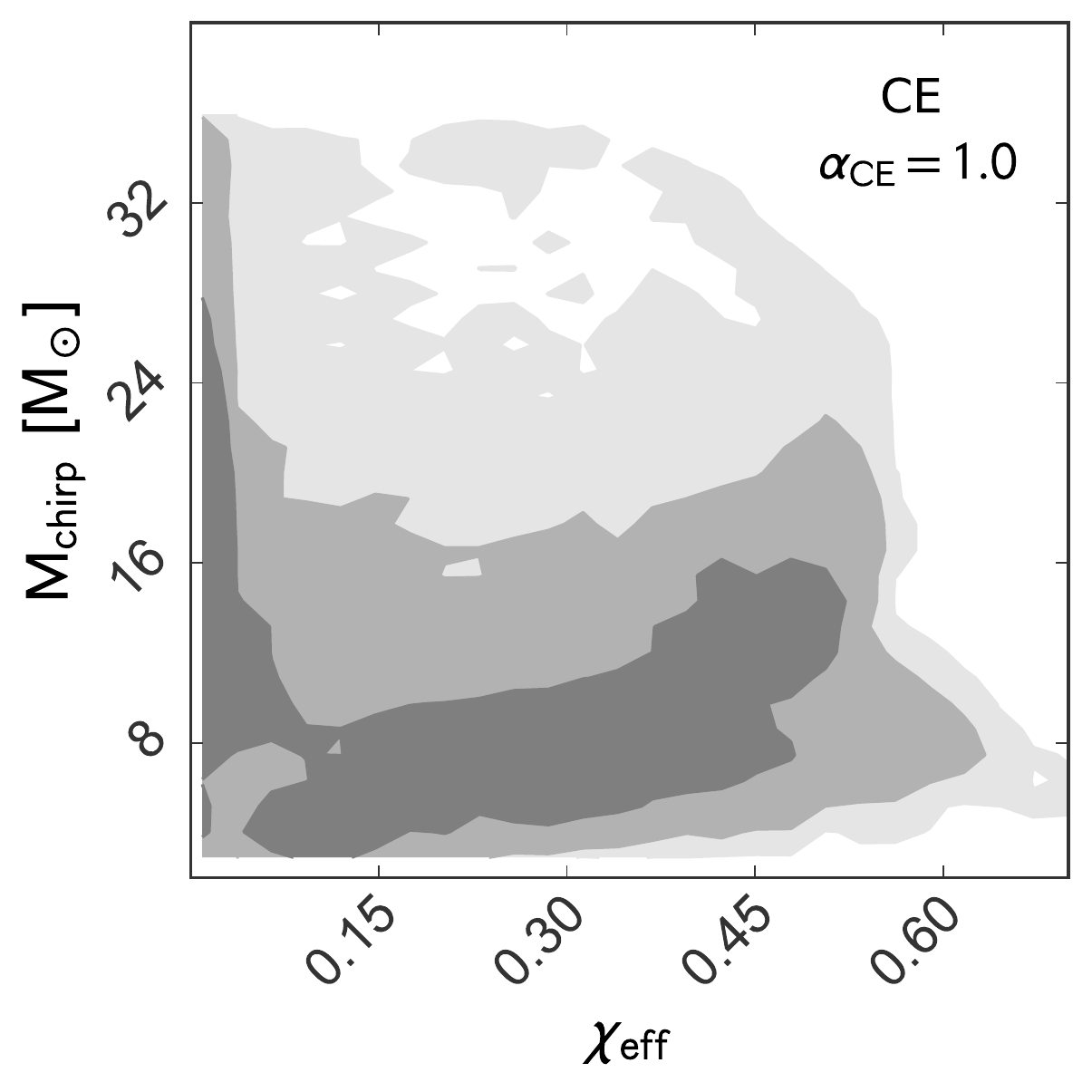}
    \end{minipage}
    \begin{minipage}[b]{0.33\textwidth}
    \includegraphics[width=\textwidth]{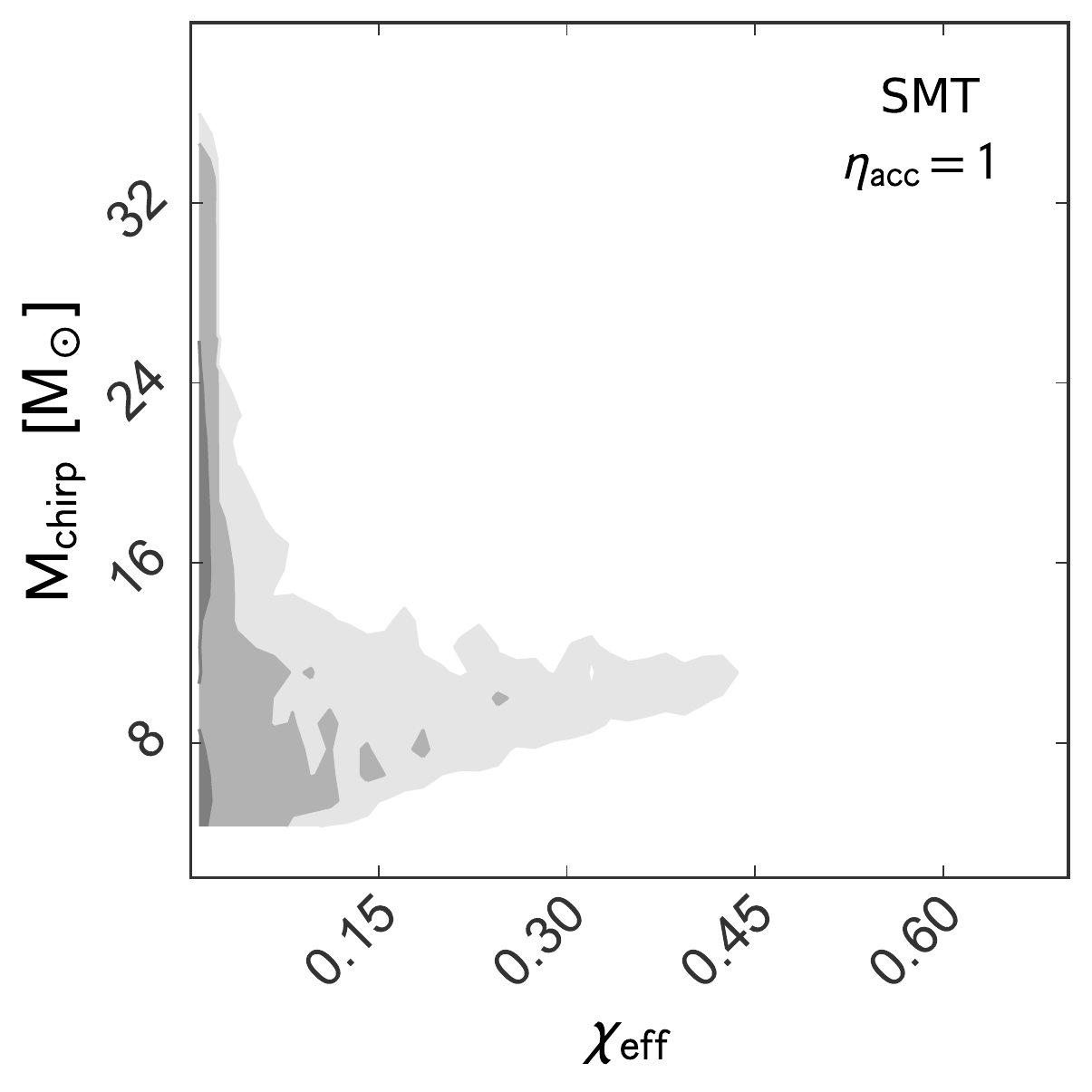}
    \end{minipage} \\  
    \begin{minipage}[b]{0.33\textwidth}
    \includegraphics[width=\textwidth]{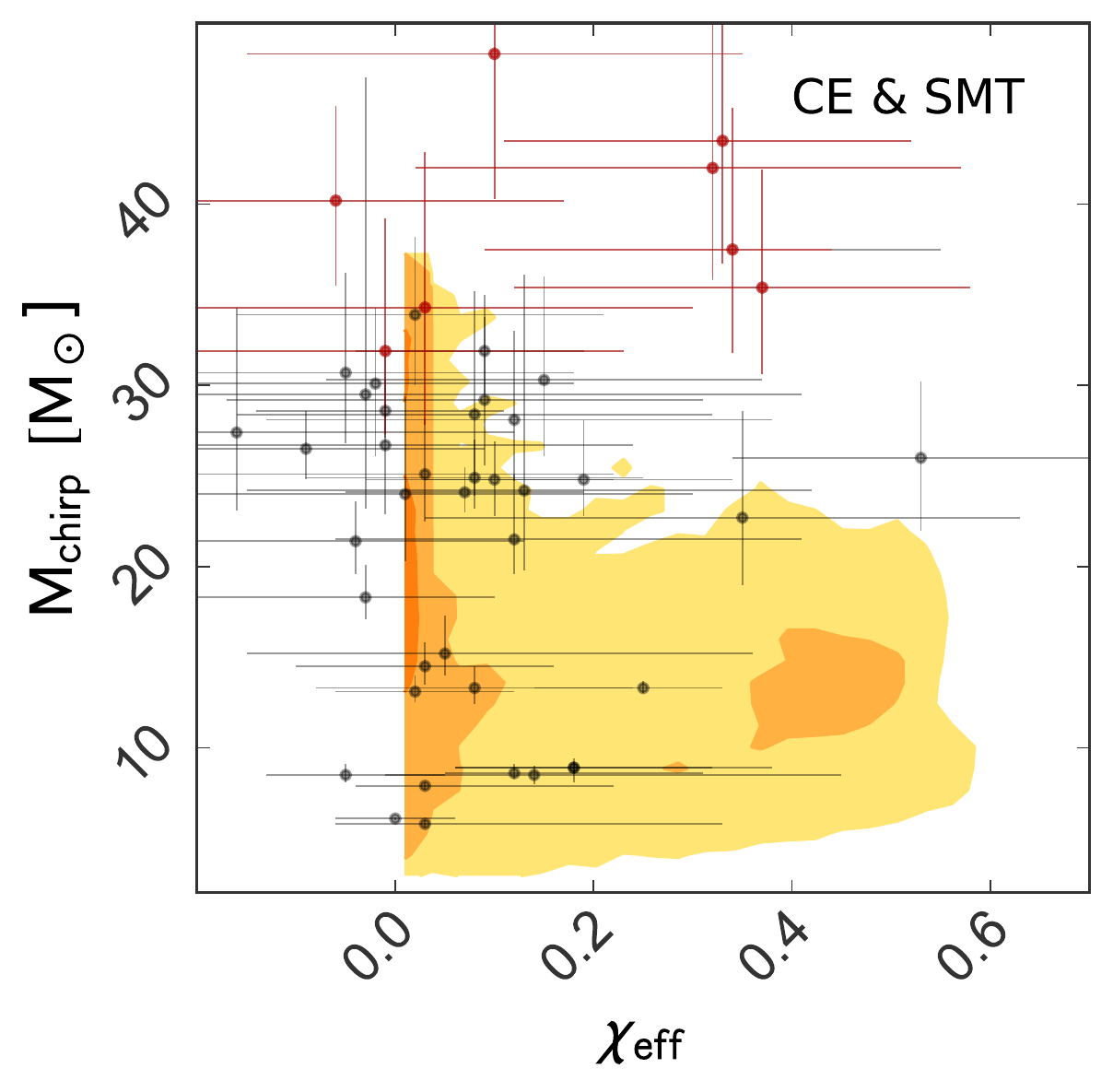}
    \end{minipage}
    \begin{minipage}[b]{0.33\textwidth}
    \includegraphics[width=\textwidth]{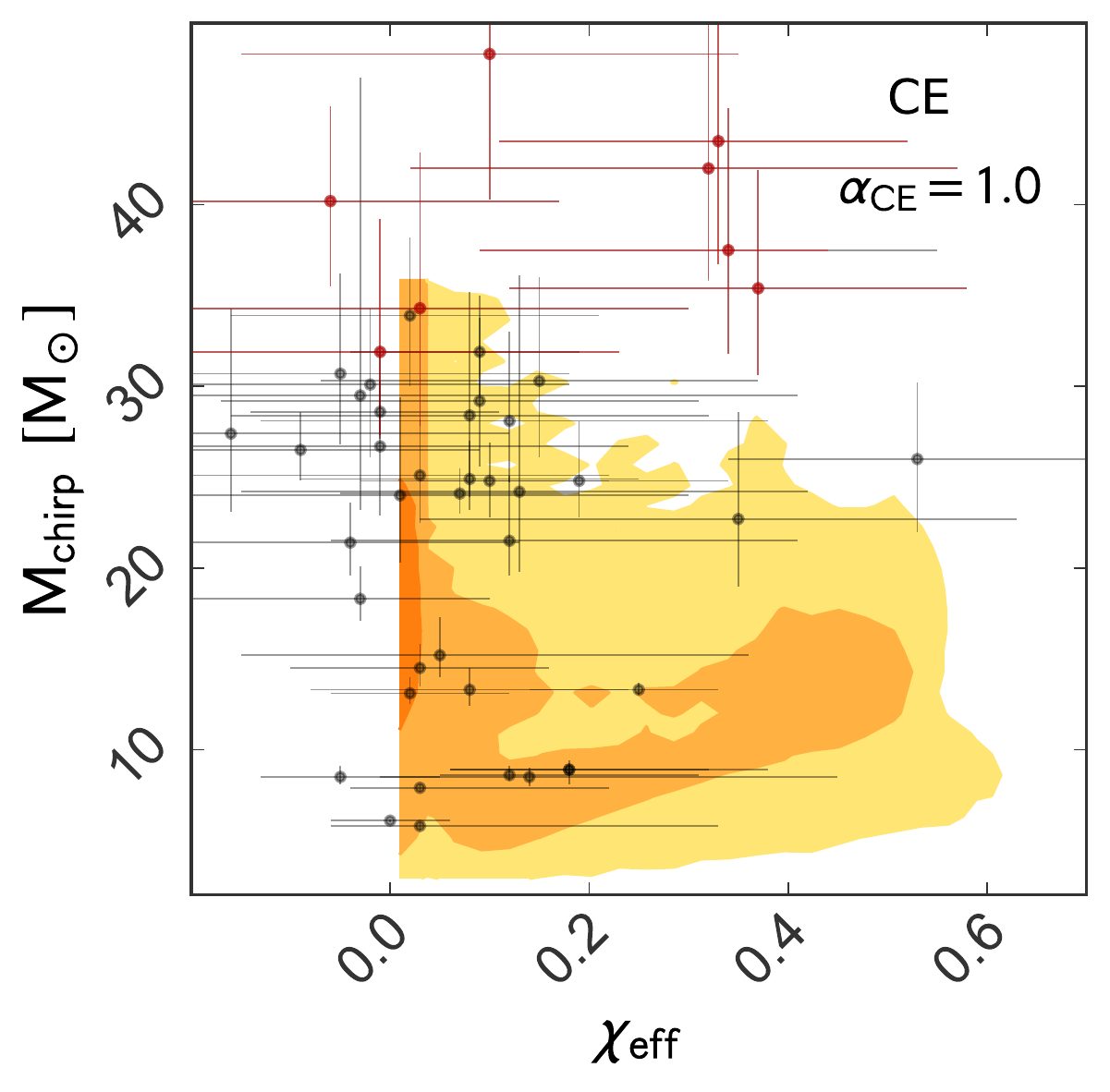}
    \end{minipage}
    \begin{minipage}[b]{0.33\textwidth}
    \includegraphics[width=\textwidth]{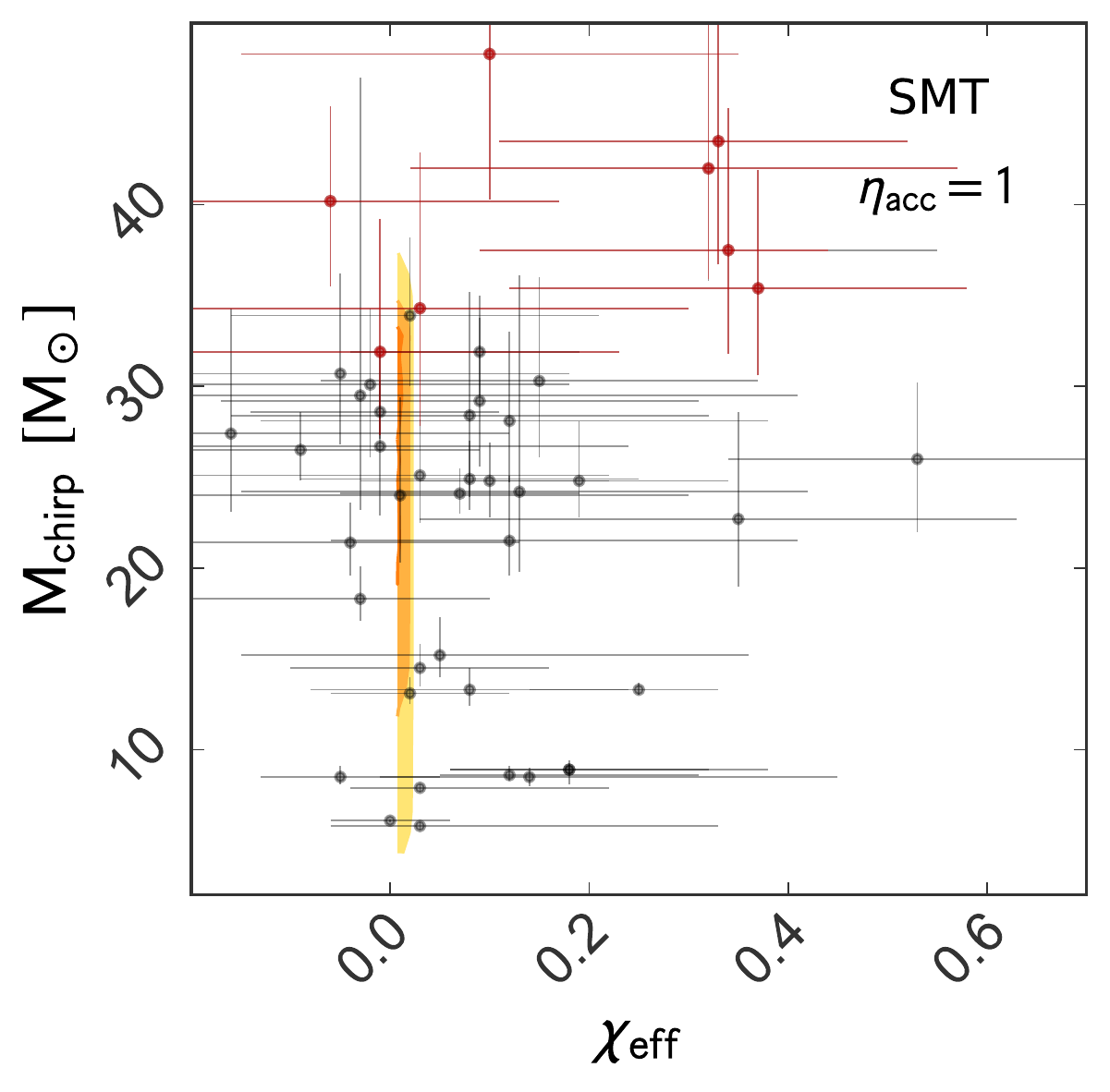}
    \end{minipage} \\
   \caption{Model predictions for the underlying (intrinsic) BBH population (grey) and the O3 detected BBH population (orange) for our reference model with $\alpha_\mathrm{CE}=1.0$ and $\eta_\mathrm{acc} = 1$. We show the joint distributions of chirp mass \mchirp and effective inspiral spin parameter \chieff for the combined CE and SMT channels (Left), CE formed BBHs (Centre) and SMT formed BBHs (Right). Lighter colours represent larger contour levels of $68\%$, $95\%$ and $99\%$, respectively, constructed with \texttt{pygtc} module \citep{Bocquet2016}. All histograms are plotted with 30 bins in the same range without any bin smoothing. We overlaid in grey the O1, O2 and O3a LVC GWTC-2 data with their $90\%$ credible intervals. The 9 events of GWTC-2 in tension with our models are indicated in red (see Sec.~\ref{sec:evidence4CH}), GW190521 is outside the plotted window.}
    \label{fig:fiducial-model}
    \end{figure*}
    
\begin{table*}
\centering                          
\begin{tabular}{c c c c c c c c c c c }        
\hline\hline                 
   &  & & Rate density & Detection rate & \multicolumn{2}{c}{\mchirp [M$_\odot$]} & \multicolumn{2}{c}{\chieff} & \multicolumn{2}{c}{$q$}\\    
   channel & $\alpha_\mathrm{CE}$ & $\eta_\mathrm{acc}$ & at $z=0.01$ & O3 sensitivity & intrinsic & detected & intrinsic & detected & intrinsic & detected \\    
   &  & & [Gpc$^{-3}$ yr$^{-1}]$ & [yr$^{-1}$] & pop. & pop. & pop. & pop. & pop. & pop.   \\    
   \hline \rule{0pt}{2.6ex}                 
   CE & 0.20 & 1 & 113.0 & 412 & $15.8^{+8.7}_{-8.9}$ & $18.5^{+8.6}_{-5.5}$ & $0.38^{+0.13}_{-0.38}$ & $0.00^{+0.42}_{-0.00}$ & $0.92^{+0.07}_{-0.18}$ & $0.95^{+0.04}_{-0.14}$ \\[1mm]
   CE & 0.35 & 1 & 17.2  & 56  & $14.7^{+11.0}_{-7.8}$ & $18.1^{+14.2}_{-8.7}$ & $0.37^{+0.17}_{-0.29}$ & $0.32^{+0.18}_{-0.32}$ & $0.84^{+0.12}_{-0.19}$ & $0.84^{+0.13}_{-0.15}$ \\[1mm] 
   CE & 0.50 & 1 & 20.4  & 61  & $13.7^{+11.5}_{-7.6}$ & $18.1^{+13.7}_{-8.8}$ & $0.27^{+0.26}_{-0.26}$ & $0.09^{+0.39}_{-0.09}$ & $0.81^{+0.13}_{-0.20}$ & $0.82^{+0.14}_{-0.15}$ \\[1mm] 
   CE & 0.75 & 1 & 29.6  & 92  & $12.5^{+11.8}_{-7.4}$ & $18.9^{+13.4}_{-9.2}$ & $0.18^{+0.33}_{-0.18}$ & $0.00^{+0.40}_{-0.00}$ & $0.80^{+0.14}_{-0.22}$ & $0.82^{+0.14}_{-0.14}$ \\[1mm]
   \textbf{CE} & \textbf{1.00} & \textbf{1} & \textbf{42.6}  & \textbf{108} & $11.0^{+12.1}_{-6.6}$ & $17.9^{+15.5}_{-8.1}$ & $0.19^{+0.32}_{-0.18}$ & $0.00^{+0.30}_{-0.00}$ & $0.79^{+0.14}_{-0.24}$ & $0.82^{+0.14}_{-0.14}$ \\[1mm]
   CE & 2.00 & 1 & 35.0  & 47  & $7.6^{+10.4}_{-4.1}$ & $16.0^{+14.8}_{-9.1}$ & $0.21^{+0.30}_{-0.21}$ & $0.00^{+0.24}_{-0.00}$ & $0.76^{+0.16}_{-0.27}$ & $0.79^{+0.17}_{-0.21}$  \\[1mm]
   CE & 5.00 & 1 & 22.7  & 15  & $5.9^{+9.5}_{-2.8}$ & $12.7^{+8.6}_{-8.0}$ & $0.16^{+0.35}_{-0.16}$ & $0.00^{+0.13}_{-0.00}$ & $0.69^{+0.22}_{-0.26}$ & $0.80^{+0.16}_{-0.29}$  \\[1mm]
   \textbf{SMT} & - & \textbf{1}  & \textbf{24.6}  & \textbf{86} & $15.3^{+14.2}_{-10.4}$ & $24.5^{+8.4}_{-11.2}$ & $0.00^{+0.07}_{-0.00}$ & $0.00^{+0.00}_{-0.00}$ & $0.72^{+0.20}_{-0.29}$ & $0.74^{+0.12}_{-0.07}$ \\[1mm]
   \hline \rule{0pt}{2.6ex}  
   CE & 1.00 & $10^3$ & 43.1 & 108 & $10.9^{+12.0}_{-6.6}$  & $17.8^{+15.4}_{-8.1}$ & $0.18^{+0.29}_{-0.18}$ & $0.00^{+0.29}_{-0.00}$ & $0.79^{+0.17}_{-0.30}$ & $0.82^{+0.14}_{-0.14}$  \\[1mm]
   SMT & - & $10^3$ & 23.5 & 85 & $15.5^{+14.2}_{-10.6}$ & $24.8^{+8.2}_{-11.3}$ & $0.01^{+0.03}_{-0.00}$  & $0.01^{+0.00}_{-0.00}$ & $0.72^{+0.19}_{-0.30}$ & $0.75^{+0.12}_{-0.07}$ \\[1mm]
   CE & 1.00 & $10^5$ & 40.8 & 97 & $10.3^{+12.0}_{-6.0}$ & $17.6^{+15.3}_{-8.1}$ & $0.18^{+0.29}_{-0.18}$  & $0.00^{+0.26}_{-0.00}$ & $0.79^{+0.17}_{-0.31}$  & $0.82^{+0.14}_{-0.15}$  \\[1mm]
   SMT & - & $10^5$ & 3.7 & 13 & $7.3^{+22.1}_{-2.0}$ & $29.4^{+6.6}_{-16.4}$ & $0.53^{+0.21}_{-0.27}$  & $0.32^{+0.13}_{-0.04}$ & $0.35^{+0.54}_{-0.16}$ & $0.89^{+0.09}_{-0.13}$ \\[1mm]
   CE & 1.00 & $10^9$ & 38.6 & 93 & $10.3^{+11.9}_{-6.0}$ & $17.4^{+15.6}_{-7.9}$ & $0.18^{+0.29}_{-0.18}$ & $0.00^{+0.26}_{-0.00}$ & $0.79^{+0.17}_{-0.31}$ & $0.83^{+0.14}_{-0.15}$  \\[1mm]
   SMT & - & $10^9$ & 0.2 & 0.3 & $8.5^{+3.2}_{-1.6}$ & $13.9^{+6.8}_{-6.0}$ & $0.78^{+0.08}_{-1.15}$ & $0.78^{+0.08}_{-0.16}$ & $0.14^{+0.07}_{-0.03}$ & $0.24^{+0.15}_{-0.11}$ \\[1mm]
\hline\hline \\
\end{tabular}
\caption{This table summarises the results of our different models. The columns report the model's channel, either CE or SMT, CE efficiency $\alpha_\mathrm{CE}$, the mass-transfer accretion efficiency limit onto compact objects $\eta_\mathrm{acc}$ in units of Eddington-limit, the local rate density at $z=0.01$ (in \rdunits), the detection rate (in yr$^{-1}$) and the median chirp mass \mchirp, effective inspiral spin parameter \chieff and binary mass ratio $q$ with their $90\%$ CI for the intrinsic (underlying) and observed BBH populations. Bold text is used to indicate our reference model.} 
\label{tab:results}
\end{table*}

\begin{table*}
\centering                          
\begin{tabular}{c c c c c c c c c c}        
\hline\hline                 
    & & \multicolumn{2}{c}{ $\chieff > 0.1$ \& $\mchirp > 15 \, \mathrm{M_\odot}$} & \multicolumn{2}{c}{$\chieff > 0.1$ \& $\mchirp < 15 \, \mathrm{M_\odot}$} & \multicolumn{2}{c}{$\mchirp > 15 \, \mathrm{M_\odot}$} & \multicolumn{2}{c}{$q > 0.8$} \\  
    \multicolumn{2}{c}{CE + SMT} & intrinsic & detected & intrinsic & detected & intrinsic & detected & intrinsic & detected \\  
   $\alpha_\mathrm{CE}$ & $\eta_\mathrm{acc}$ & pop. & pop. & pop. & pop. & pop. & pop. & pop. & pop. \\  
   \hline \rule{0pt}{2.6ex}              
   0.20 & 1 & 0.33 & 0.05 & 0.27 & 0.01 & 0.56 & 0.88 & 0.77 & 0.83 \\
   0.35 & 1 & 0.38 & 0.22 & 0.44 & 0.08 & 0.48 & 0.83 & 0.61 & 0.40 \\
   0.50 & 1 & 0.22 & 0.12 & 0.46 & 0.08 & 0.42 & 0.82 & 0.52 & 0.37 \\
   0.75 & 1 & 0.11 & 0.04 & 0.46 & 0.05 & 0.34 & 0.83 & 0.47 & 0.41 \\
   1.00 & 1 & 0.09 & 0.02 & 0.49 & 0.03 & 0.27 & 0.80 & 0.45 & 0.42 \\
   2.00 & 1 & 0.06 & 0.01 & 0.59 & 0.03 & 0.15 & 0.78 & 0.39 & 0.29 \\
   5.00 & 1 & 0.01 & 0.00 & 0.53 & 0.01 & 0.13 & 0.82 & 0.28 & 0.24 \\
   \hline \rule{0pt}{2.6ex}
   1.00 & $10^3$ & 0.08 & 0.02 & 0.50 & 0.03 & 0.27 & 0.81 & 0.45 & 0.43 \\
   1.00 & $10^5$ & 0.07 & 0.13 & 0.55 & 0.06 & 0.22 & 0.74 & 0.47 & 0.68 \\
   1.00 & $10^9$ & 0.07 & 0.02 & 0.55 & 0.06 & 0.22 & 0.71 & 0.48 & 0.65 \\
\hline\hline \\
\end{tabular}
\caption{This table summarises the relative fractions of BBHs formed through CE and SMT channels combined for some arbitrary parameter space divisions (according to the column labels) for both the intrinsic (underlying) and observed BBH populations.  The columns report the CE efficiency $\alpha_\mathrm{CE}$, mass-transfer accretion efficiency limit onto compact objects $\eta_\mathrm{acc}$ in units of Eddington-limit and the relative fraction of events in the parameter space slices: $\chieff > 0.1$ \& $\mchirp > 15 \, \mathrm{M}_\odot$, $\chieff > 0.1$ \& $\mchirp < 15 \, \mathrm{M}_\odot$, $\mchirp > 15 \, \mathrm{M}_\odot$, and $q > 0.8$.} 
\label{tab:relative_fractins}
\end{table*}

\section{Results}\label{sec:results} 

We use our models to predict the distributions of some of the main GW observables: the effective inspiral spin parameter \chieff, the chirp mass \mchirp and the binary mass ratio $q$. We investigate how these distributions change for different CE and accretion efficiencies. The distributions for SMT and CE channels are obtained by distributing the synthetic BBH population across the cosmic history of the Universe as described in Sec.~\ref{sec:weights}.

Our models use detailed binary evolution simulations to determine the spin of the second-born BH, assuming that the first-born BH is formed with a negligible spin $a_1 \simeq 0$ because of the assumed efficient AM transport \citep{2018A&A...616A..28Q,2019ApJ...881L...1F}. If the second MT is stable the first-born BH can accrete material and spin up \citep{1974ApJ...191..507T}. Nevertheless, because in our reference models we assume Eddington-limited accretion efficiency onto compact objects, the accreted mass is small; this leads to small $a_1 \simeq 0$ also for the SMT channel. The Eddington limited accretion onto the BH is a crucial assumption for the existence of this channel. In Sec.~\ref{sec:SMT} we show that if highly super-Eddington accretion onto the BH is allowed, the SMT channel contributes to a negligible part to the BBH rate density compared to the CE channel.

\begin{figure}
    \includegraphics[width=\columnwidth]{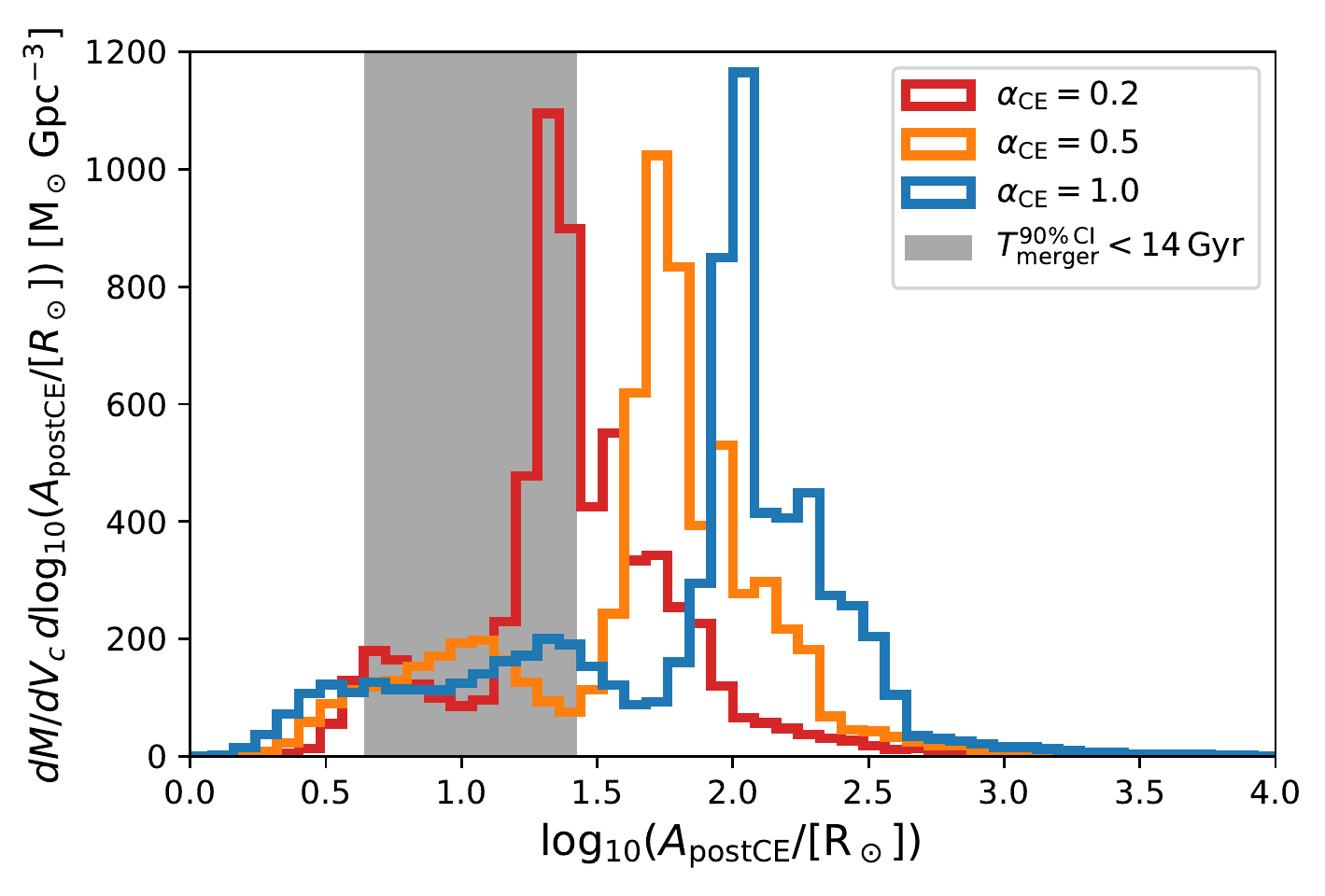}
     \caption{Orbital separation of BH--He-star binaries post CE for $\alpha_\mathrm{CE} \in [0.2,0.5,1.0]$ represented with solid lines according to the legend. The histogram has units of $\mathrm{M}_\odot \, \mathrm{Gpc}^{-3}$ and accounts for the total stellar mass formed per comoving volume integrated over the Universe cosmic history per log-orbital period bin. The grey shaded area represent the $90\%$ CI of the systems forming merging BBHs with $T_\mathrm{merger} < 14~\mathrm{Gyr}$ for the $\alpha_\mathrm{CE} = 0.2$ model (the other models have similar CIs). As the CE efficiency is lowered, the orbital separations become smaller and the distributions move to the left: for $\alpha_\mathrm{CE}=0.2$ the orbital separation peak enters the grey area boosting the merger rate.
     }
     \label{fig:ppostCE}
\end{figure}

\begin{figure}
    \includegraphics[width=\columnwidth]{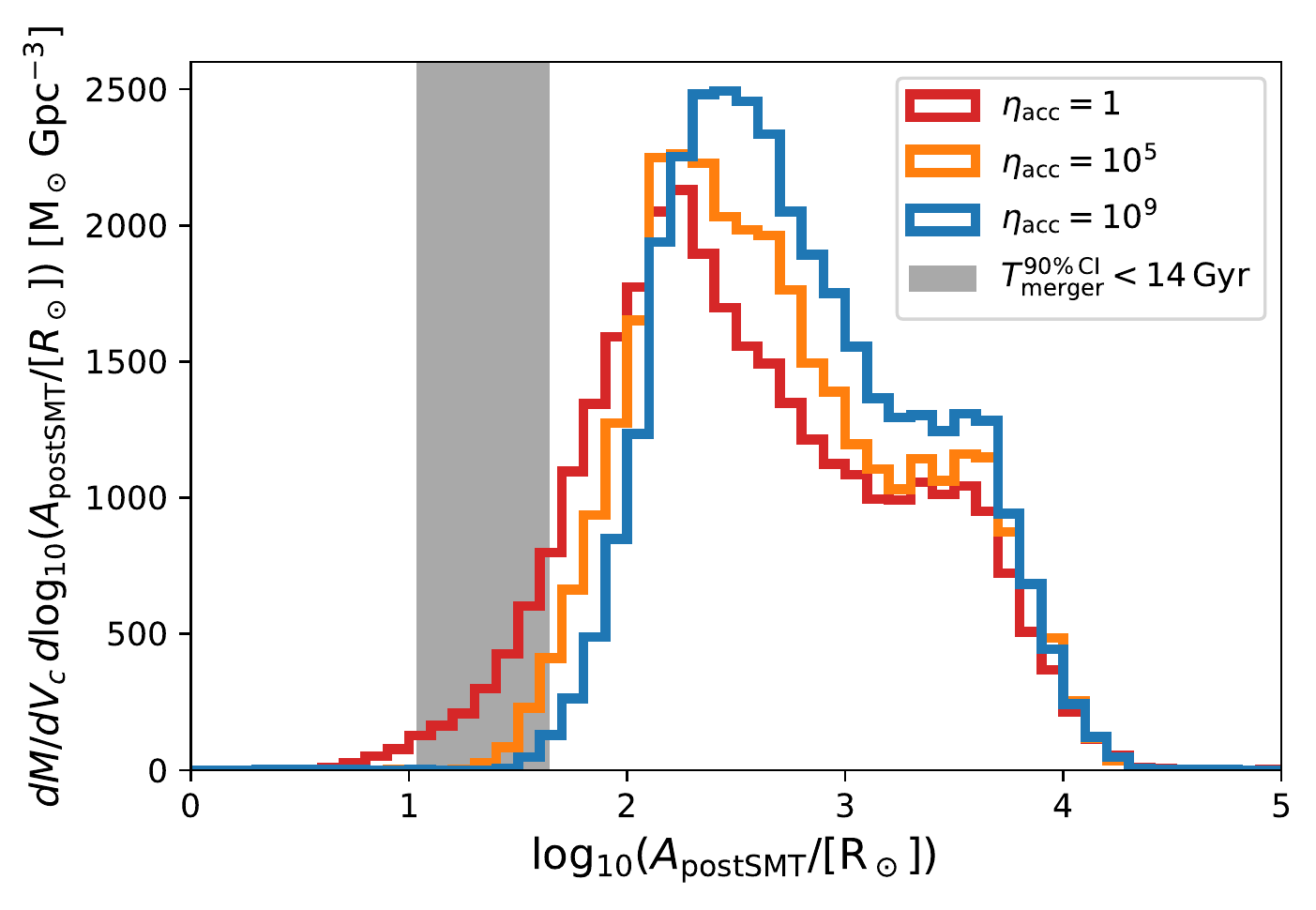}
     \caption{Orbital separation of BH--He-star binaries post SMT for accretion efficiency limit $\eta_\mathrm{acc} \in [1,10^5,10^9]$ (in units of Eddington-limit) represented with solid lines according to the legend. The histogram has units of $\mathrm{M}_\odot \, \mathrm{Gpc}^{-3}$ and accounts for the total stellar mass formed per comoving volume integrated over the Universe cosmic history per log-orbital period bin. The grey shaded area represent the $90\%$ CI of the systems forming merging BBHs with $T_\mathrm{merger} < 14~\mathrm{Gyr}$ for the model with $\eta_\mathrm{acc} = 1$ (the other models have similar CIs). As the accretion efficiency is increased, the orbital separations become larger and the distributions move to the right: for highly super-Eddington accretion efficiency $\eta_\mathrm{acc}=10^9$ the tail almost exits the grey area, hence, decreasing the merger rate.
     }
     \label{fig:ppostSMT}
\end{figure}

\subsection{Underlying BBH population}

\subsubsection{Common envelope channel}\label{sec:CE}

In our CE channel, only the spin of the second-born BH is non-zero and, hence, contributes to the \chieff parameter. The BH progenitor can be tidally spun up during binary evolution after the CE event. The efficiency of tides depends strongly on the orbital separation as the synchronisation timescale $T_\mathrm{syn} \propto A^{17/2}$, see Eq.~\eqref{eq:tides}. On the other hand, stellar winds can cause the binary to lose mass, widening the orbit and reducing or neutralising the effects of tides. He-stars have wind mass loss rates that strongly depend on metallicity. Hence, a second-born, tidally spun-up BH can only occur at low metallicity as shown in our detailed binary simulation (see Fig.~\ref{Fig:grids-bgi}). A key point is that our detailed simulations do not show a dichotomy between tidally locked and non-spinning second-born BHs, but smoothly cover the whole range of $a_2 \in [0,1]$, (e.g. top panel of Fig.~\ref{fig:grids-small}). This was pointed out by \citet{2018A&A...616A..28Q} and \citet{2020A&A...635A..97B} and is in contrast with results from semi-analytical models \citep{2017ApJ...842..111H,2018MNRAS.473.4174Z,2018PhRvD..98h4036G}. 

In the top panels of Fig.~\ref{fig:fiducial-model} we show the joint distribution of \chieff and \mchirp for the underlying BBH population of the CE channel for the reference model with $\alpha_\mathrm{CE} = 1$ alongside the SMT channel and their combination. For the CE channel, we can see that the underlying BBH population has a non-negligible amount of positive \chieff mergers due to the tidal spin-up of the second born BH's progenitor. This is in agreement with our previous models \citep{2020A&A...635A..97B} obtained using the \compas{} code \citep{2017NatCo...814906S,2018MNRAS.481.4009V} which is based on the same stellar model fits but implements binary interactions differently (a comparison between the two codes is beyond the scope of this work). Because of the anti-correlation between the merger timescale $T_\mathrm{merger}$ and \chieff (Sec.~\ref{sec:BBHpop}), these highly spinning systems merge soon after their formation. Current GW detectors probe only small redshifts ($z \lesssim 1$), well below the peak of the cosmic SFR ($z \sim 2$) where most of these systems are created and merge, as low metallicity environments are required for efficient tidal spin up. 

CE efficiency has an important role in the determination of the post-CE orbital distribution. This is because  $\alpha_\mathrm{CE}$ correlates approximately linearly with the post-CE orbital separation $A_\mathrm{postCE}$, see Eq.~\eqref{eq:CE}. Models assuming inefficient CE ejection ($\alpha_\mathrm{CE} < 1.0$) result in tighter post-CE orbits and more systems merging during CE compared to larger $\alpha_\mathrm{CE}$ values (see Fig.~\ref{fig:ppostCE}). This occurs because the binaries need to deposit more orbital energy into the envelopes to successfully eject them. The opposite is true when assuming an efficient CE ejection ($\alpha_\mathrm{CE} > 1.0$). We expect, on average, larger \chieff values for models with lower CE efficiency parameters, as more systems will undergo tidal spin up and small \chieff for models assuming ultra-efficient CE ejection, as tides are weaker at larger orbital separations. Indeed, this trend is what we find. In Table~\ref{tab:results}, we report the median \chieff, \mchirp and $q$ with their $90\%$ confidence interval (CI) for our different CE efficiencies models. For the underlying (intrinsic) BBH population, we observe a monotonic decrease of all these quantities for increasing $\alpha_\mathrm{CE}$ (from $0.2$ to $5.0$). This trend is also found if we look at the relative fractions of massive and highly spinning BBHs, namely with $\chieff > 0.1$ and $\mchirp > 15 \, \mathrm{M}_\odot$, in Table~\ref{tab:relative_fractins}. In both tables we see that on average models with small $\alpha_\mathrm{CE}$ have larger \mchirp and $q$. This is because for the same orbital separation, massive binaries have a larger orbital energy reservoir compared to lighter systems and, hence, can deposit more energy into the envelope without shrinking to the point where they merge in the CE phase. 

In Table~\ref{tab:results} we also report the local rate density for the cosmic time bin centreed at $z=0.01$. The reference model with $\alpha_\mathrm{CE} = 1.0$ has a local rate density of $42.6~\rdunits$. If we increase $\alpha_\mathrm{CE}$, the post-CE orbital separations are larger, hence, the rate density decreases because fewer systems merge within the Hubble time. On the other hand, if we decrease $\alpha_\mathrm{CE}$, more systems merge during the CE event and the rate density decreases as well. This trend is not followed by the model with $\alpha_\mathrm{CE}=0.2$ where the rate density suddenly jumps up to 113.0~\rdunits. To understand the sudden increase in the rate density of this model, we need to carefully look at the post-CE binary orbital separations. In Fig.~\ref{fig:ppostCE} we show a histogram of all BH--He-star orbital separations surviving CE for $\alpha_\mathrm{CE} \in [0.2,0.5,1.0]$ (solid lines). The synthetic BBH population is weighted according to Eq.~(B.10) of \citet{2020A&A...635A..97B} which integrates the redshift- and metallicity-dependent SFR across the cosmic history of the Universe. In grey, we show the 90\% CI of systems forming merging BBHs with $T_\mathrm{merger} < 14~\mathrm{Gyr}$ for the model with $\alpha_\mathrm{CE} = 0.2$ (the other models have similar CIs). Systems with orbital separations smaller than the left boundary of the CI either form BH-NS binaries, merge during the BH--He-star evolution or widen the orbits (because of wind driven mass loss rate) past the point where they will merge within the Hubble time. Systems on the right of the CI form double compact objects with merging timescales larger than the Hubble time. In this figure we see that as the CE efficiency decreases, the orbital separations decrease. The total orbital separation distributions present a large peak of orbital separation preceded by a smaller flatter distribution of orbital separation. For $\alpha_\mathrm{CE}=0.2$ we see that this large peak of orbital separation enters the merging BBH population (grey area of Fig.~\ref{fig:ppostCE}). This is the source of the sudden increase of the rate density. 

The peak of BH--He-star orbital separations post CE is a metallicity product. All binaries going through CE are evolving during the He-burning phase. In our models, systems that are in the HG at onset of CE are considered to merge during the CE \citep[because we assume the pessimistic CE scenario; see][]{2007ApJ...662..504B}. The maximum stellar radius of a star in the HG is metallicity dependent. Even though, on average stars with high metallicity have larger radii during this phase compared to lower metallicity stars, they have similar super-giant phase radii, see for example, Fig.~7 of \citet{2010ApJ...725.1984L}. This implies that binaries with high metallicities sample a smaller range of orbital separations at onset of CE, with the donor star having passed the HG phase, compared to binaries with lower metallicities. Therefore, low metallicity BH--He-star binaries sample a wide (approximately flat) $A_\mathrm{preCE}$ distribution which result to a wide (also approximately flat) range of $A_\mathrm{postCE}$, while high metallicity systems sample a narrow $A_\mathrm{preCE}$ distribution which result in a narrow $A_\mathrm{postCE}$ range. Fig.~\ref{fig:ppostCE} shows the combination of these distributions for all metallicities. Since the average metallicity in the Universe is a monotonically increasing function, the yield of binaries at low metallicities is smaller than the yield at larger metallicities, hence the larger $A_\mathrm{postCE}$ peak.

\subsubsection{Stable mass transfer channel}\label{sec:SMT}

In our SMT channel, the spin of both BHs can be non-zero and hence affect the \chieff parameter. Since in our reference model we assume Eddington-limited MT, the amount of accreted material is negligible compared to the BH mass and leads to small spins ($a_1 \simeq 0.002$ is the largest value in our population). In Table~\ref{tab:results} we show that models with super-Eddington accretion efficiency limits result in larger first-born BH spins, as the BH is allowed to accrete at highly super-Eddington rates, and, hence, result in larger median \chieff. After the second MT phase, tides can further spin up the second-born BH progenitor if the orbits are tight enough (this requires $p < 1$ day). Since the orbits cannot shrink as efficiently as in the CE channel, most of the systems formed through this evolutionary path will not undergo tidal spin up. Since the CE efficiency does not affect this evolutionary path, here we report only values from the model with $\alpha_\mathrm{CE} = 1.0$.

In Fig.~\ref{fig:fiducial-model} we show the underlying joint distribution of \chieff and \mchirp for the SMT channel alongside the CE channel and their combination. We see that the underlying SMT BBH population presents a non-zero \chieff contour at the 95\% level. These are the systems accreting during the second MT event with tidal spin up during the subsequent phase. Even though the non-zero \chieff distribution is a small part of the overall population (cf.\ median \chieff in Table~\ref{tab:results}) this subpopulation has an astrophysical consequence. During the second MT, the accreting BHs are thought to form an accretion disk with strong X-ray emission. This partly explains the high-end of the luminosity function of stellar X-ray sources in galaxies \citep[e.g. ultraluminous X-ray sources;][]{2002ApJ...568L..97B,2004ApJS..154..519S,2020MNRAS.498.4790K}.

In Table~\ref{tab:results} we report the rate density contribution of SMT channel to be $24.6~\rdunits$. This value is comparable to the contribution of the CE channel for $\alpha_\mathrm{CE} \in [0.35,5.0]$. This result is consistent with other studies \citep[e.g.][]{2019MNRAS.490.3740N} and is strongly dependent on the assumed accretion efficiency limit onto the BH. If we allow for super-Eddington accretion the BBH rate density will decrease. The drop in the rate density when allowing for super-Eddington accretion occurs because conservative MT does not shrink the orbit as efficiently as non-conservative MT (in the case of Eddington-limited accretion) and thus the BBHs formed post-MT are more commonly too wide to merge within the Hubble time. This can be seen in Fig.~\ref{fig:ppostSMT} where we show a histogram of all BH--He-star orbital separation post SMT for different accretion efficiency limits $\eta_\mathrm{acc} \in [1, 10^5, 10^9]$ (in units of Eddington-limit), solid lines, and the 90\% CI of systems forming merging BBHs with $T_\mathrm{merger} < 14$ Gyr, in grey. For not Eddington-limited accretion  (here arbitrarily limited to up to $10^9$ times the Eddington-limit) the rate density contribution of this channels drops by two orders of magnitudes down to $0.2~\rdunits$ while just decreasing the CE channel rate by $\sim 10\%$. This is a negligible contribution to the BBH merger rate compared to the yield of the CE channel. 

When allowing for super-Eddington accretion, the first-born BH accretes a non-negligible amount of matter leading to a different mass radio distribution compared to Eddington-limited SMT models. In Table~\ref{tab:results} we can see that the median mass ratio decreases from $0.72$ to $0.35$ and $0.14$ for increasing $\eta_\mathrm{acc}$. Small mass ratios are also found by the \texttt{BPASS} team \citep{2017PASA...34...58E} who, in their models, allow for super-Eddington accretion onto BHs. As mentioned above, the first-born BH is spun-up when mass is accreted onto it. Indeed, for increasing $\eta_\mathrm{acc}$, our SMT models predict larger median \chieff.

\begin{figure*}
    \centering
    \includegraphics[width=\textwidth]{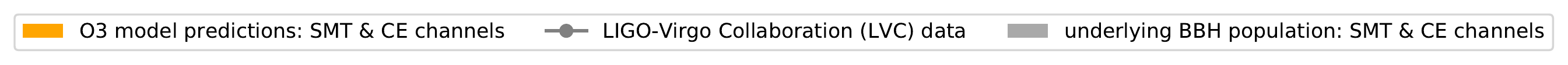} \\
    \includegraphics[width=\textwidth]{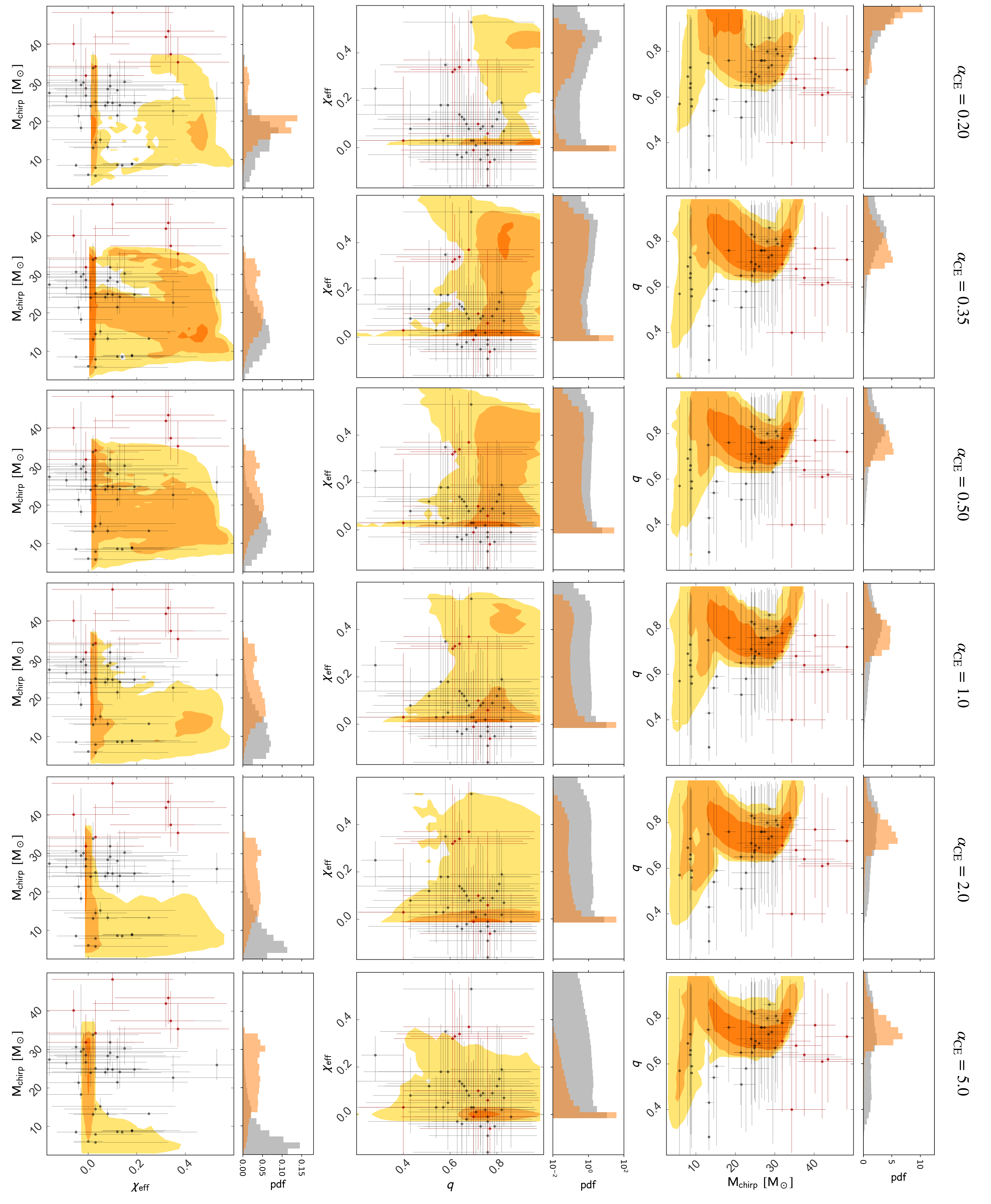}
     \caption{Model predictions for the O3 detected BBH population of CE and SMT channels combined for $\alpha_\mathrm{CE} \in [0.2, 0.35, 0.5, 1.0, 2.0, 5.0]$ and $\eta_\mathrm{acc}=1$, in orange. We show the joint distributions of chirp mass $M_\mathrm{chirp}$, effective inspiral spin parameter $\chi_\mathrm{eff}$ and binary mass ratio $q$. Lighter colours represent larger contour levels of $68\%$, $95\%$ and $99\%$, respectively, constructed with \texttt{pygtc} \citep{Bocquet2016}. All histograms are plotted with 30 bins without any bin smoothing. We overlaid in grey the LVC GWTC-2 data with their 90\% credible intervals. The 9 events of GWTC-2 in tension with our models are indicated in red (see Sec.~\ref{sec:evidence4CH}). We also show the one-dimensional projection of each quantity overplotted on the underlying (intrinsic) BBH population, in grey. For visualisation purposes, the $\chi_\mathrm{eff}$ probability density function (pdf) is plotted in log-scale. For a visualisation of each model separately see Appendix~\ref{app:model-results}.
     }
     \label{fig:models}
\end{figure*}   

\subsection{Detected BBH population}

O3a lasted $6$ months and saw the detection of $37$ GW signals from BBH mergers resulting in a total of 177.3 days of data suitable for coincident analysis. These detections translate to a detection rate of $76~\mathrm{yr}^{-1}$. In our model comparison with the data we include also the BBH detections of the first two observing runs O1 and O2 \citep{2019PhRvX...9c1040A} for a total of 47 events. Although evidence for additional signals in O1 and O2 data has been presented by other groups \citep[e.g.][]{2019arXiv191009528Z,2019PhRvD.100b3007Z,2020PhRvD.101h3030V,2020ApJ...891..123N}, we do not include them in order to simplify our analysis and only consider a single treatment of selection effects. At the same time, adding a few additional, low significance events in the observed population is not expected to add significant discriminating power. For all events, we assume simulated O3 sensitivity (mid high and late low from \citealt{2018LRR....21....3A}) as the observable distributions of \chieff and \mchirp are weakly dependent on the detector sensitivity for the channels considered \citep{2020A&A...635A..97B}.

The detected joint distributions of \chieff and \mchirp of our reference model with $\alpha_\mathrm{CE} = 1.0$ are presented in the bottom panels of Fig.~\ref{fig:fiducial-model}. The figure shows CE and SMT channels alongside their combination. For a visual comparison with the observations, we add the LVC GW detections with their $90\%$ credible intervals in grey. We can clearly see that the SMT channel only contributes with zero \chieff and large \mchirp systems to the detected population. Hence, in our model, the only source of non-zero \chieff in the detected BBH population comes from the CE channel. 

In Fig.~\ref{fig:models} we show the predicted two-dimensional distributions of \chieff, \mchirp and $q$ for the combined CE and SMT detected population at different CE efficiencies. We can clearly see how the models with inefficient CE ($\alpha_\mathrm{CE} < 1$) lead to overall larger \chieff values compared to ultra efficient CE in the detected population ($\alpha_\mathrm{CE} > 1$). Next to each panel we also include the normalised one-dimensional histogram of each quantity where we also show the underlying BBH population for comparison. The one-dimensional \mchirp histograms show how the selection effect favouring higher BH masses changes the distribution, while in the one-dimensional \chieff histograms we can see how the detectable population mostly probes low $\chi_\mathrm{eff}$. This happens because the GW detectors probes small redshifts ($z \lesssim 1$) while highly spinning systems tend to form at high redshifts ($z \sim 2$) and low metallicity environments, and merge quickly at a redshift close to the one of their formation (see the discussion about the anti-correlation between \chieff and $T_\mathrm{merger}$ in Sec.~\ref{sec:BBHpop}). These high redshifts are outside current GW detection horizons. 

In Table~\ref{tab:results} we report the detection rate of each model for O3 sensitivity. For the SMT channel our model predicts a detection rate of $86~\mathrm{yr}^{-1}$ while the detection rate for the CE model with $\alpha_\mathrm{CE}=1.0$ is $108~\mathrm{yr}^{-1}$. As we increase or decrease CE efficiency we lower the detection rate to $56~\mathrm{yr}^{-1}$ for $\alpha_\mathrm{CE} = 0.35$ and $15~\mathrm{yr}^{-1}$ for $\alpha_\mathrm{CE} = 5.0$. The model with $\alpha_\mathrm{CE} = 0.2$ overpredicts the detections with $412~\mathrm{yr}^{-1}$ (see Sec.~\ref{sec:CE} for an explanation). On the other hand, the SMT model with highly super-Eddington accretion efficiency limit predicts a detection rate of $0.2~\mathrm{yr}^{-1}$ which is negligible compared to the CE channel contribution. Within the probed mass-transfer physics uncertainties, the combination of the two channels is roughly consistent with the observed rate. While our models are consistent with observations, the event rate does depend on many other uncertain evolutionary parameters \citep[e.g.][]{2015ApJ...806..263D,2018MNRAS.474.2959G,2018MNRAS.477.4685B} and metallicity-specific star formation history \citep[e.g.][]{2019MNRAS.482.5012C,2019MNRAS.490.3740N} which we have not explored. Therefore, while our results can illustrate the expected trend when varying CE efficiency, we cannot make definitive statements on the true value of $\alpha_\mathrm{CE}$ without also considering the other evolutionary parameters.

\subsection{Evidence for additional formation channels}\label{sec:evidence4CH}

In GWTC-2 there are BBH events that appear marginally consistent or inconsistent explained by our models of isolated binary evolution through CE or SMT. 
Using individual events to discriminate between models should be done with caution, as the information that individual events carry can be strongly affected by the choice of priors used in the analysis \citep[e.g.][]{2020ApJ...895L..28M,2020ApJ...899L..17Z,2020ApJ...904L..26F}.
Instead, one should attempt to derive conclusions based on the combined detected population. 
In this section we discuss such events which may originate from other active formation scenarios (see the discussion in Sec.~\ref{sec:introduction}), while in the following section we perform a model comparison based on the combined sample of events.

In the catalogue we have two high-significance events with asymmetric masses: GW190412 and GW190814. GW190412 has a binary mass ratio of $q=0.28^{+0.13}_{-0.06}$ \citep{2020arXiv200408342T} while GW190814 has $q=0.112^{+0.008}_{-0.009}$ \citep{2020ApJ...896L..44A}. We find that these small mass ratios are consistent at the $90\%$ level of BBHs formed though SMT with highly super-Eddington accretion. In these models the first-born BH accretes material during the second MT phase leading to unequal mass ratios. However, these models predict large \chieff values as the first-born BH is spun up during accretion \citep{1974ApJ...191..507T}. The $90\%$ CI of \chieff in this model is not consistent with the $\chieff = 0.25^{+0.08}_{-0.11}$ and $\chieff = -0.002^{+0.060}_{-0.061}$ of GW190412 and GW190814, respectively. If one assumes a different, astrophysically-motivated prior, such as a prior that assumes non-spinning primary BHs (different formation channels have different priors), rather than the one used in the LVC analysis, GW190412's inferred mass ratio increases to $0.34 \leq q \leq 0.47$ at the $90\%$ level \citep{2020ApJ...899L..17Z}. The latter is marginally consistent with our models. The case of GW190814, with its lower-mass component being a $2.6\,\mathrm{M}_\odot$ compact object, remains a challenge to explain with isolated BBH formation \citep{2020ApJ...899L...1Z}.

GW190521 is a GW signal with a BBH source with high component masses, $85^{+21}_{-14} \, \mathrm{M}_\odot$ and $66^{+17}_{-18} \, \mathrm{M}_\odot$ \citep{2020PhRvL.125j1102A}. Accounting for pulsational pair instability \citep[PPI;][]{2016MNRAS.457..351Y,2017ApJ...836..244W,2019ApJ...882...36M} and a pair instability supernova  \citep[PISN;][]{1964ApJS....9..201F,1967ApJ...148..803R,1967PhRvL..18..379B} uncertainties, the primary mass falls in the mass gap predicted by PISN at around $[40$--$65, 120] \, \mathrm{M}_\odot$ \citep{2003ApJ...591..288H,2017MNRAS.470.4739S,2018MNRAS.474.2959G,2018ApJ...863..153T,2019ApJ...878...49W,2019ApJ...882...36M,2019ApJ...887...53F,2020A&A...640L..18M}. Our work adopt fits to PPI and PISN models of \citet{2019ApJ...882...36M} which for metallicity $0.1 \, Z_\odot$ find that the maximum BH mass is $\sim 44 \, \mathrm{M}_\odot$. Hence, this system is a poor fit to our CE and SMT models. This conclusion is consistent with other studies; for example, \citet{2020ApJ...897..100V} investigated isolated binary evolution with super-Eddington accretion without finding any merging BBH systems with a total mass exceeding $100 \, \mathrm{M}_\odot$.  Alternatively, \citet{2020ApJ...904L..26F} showed that if the event is analysed with a prior on the less massive BH of $m_\mathrm{BH2} < 48 \, \mathrm{M}_\odot$ at $90\%$ credibility, then, the primary BH has a $39\%$ probability of being above the PISN gap. In our models we did not explore stellar and binary evolution above the PISN gap and hence, cannot rule out the formation through CE or SMT.

Amongst the O2 detections there is one event marginally consistent with our models: GW170729 \citep{2019PhRvX...9c1040A}. This event has high chirp mass of $\mchirp = 35.4^{+6.5}_{-4.8} \, \mathrm{M}_\odot$ and a high effective inspiral spin of $\chieff = 0.37^{+0.21}_{-0.25}$ \citep{2019PhRvX...9c1040A}. Fig.~\ref{fig:models} shows that only models with low $\alpha_\mathrm{CE}$ are consistent with this event at the $99\%$ level. When we calculate the likelihood of this system originating from our CE and SMT models using Eq.~\eqref{eq:likelihood}, we find it to be extremely small compared to the other events.

Similarly, we can identify a groups of events in GWTC-2 with primary BH masses which support masses larger than our PPI maximum BH mass $\sim 44 \, \mathrm{M}_\odot$ \citep{2019ApJ...882...36M}. These are {GW190413\_134308, GW190519\_153544, GW190521,} {GW190602\_175927, GW190620\_030421, GW190701\_203306, } GW190706\_222641 and GW190929\_012149. These events, given our CE and SMT models, have extremely small likelihoods with respect to others, and hence cannot be readily explained by our models.

All the events discussed in this section, perhaps with the exception of GW190412 and GW190814, are likely not the outcome of isolated binary evolution though CE or SMT, given our models. Without considering explicitly models for alternative channels, which is outside the scope of our study, we cannot make a conclusive statement on their origins. However, if we assume that all these systems originated from a different formation channel we can put a lower bound on the contamination fraction in the detected BBH population from other channels to be $9/47 \simeq 0.2$.

\begin{figure}
\centering
\includegraphics[width=\columnwidth]{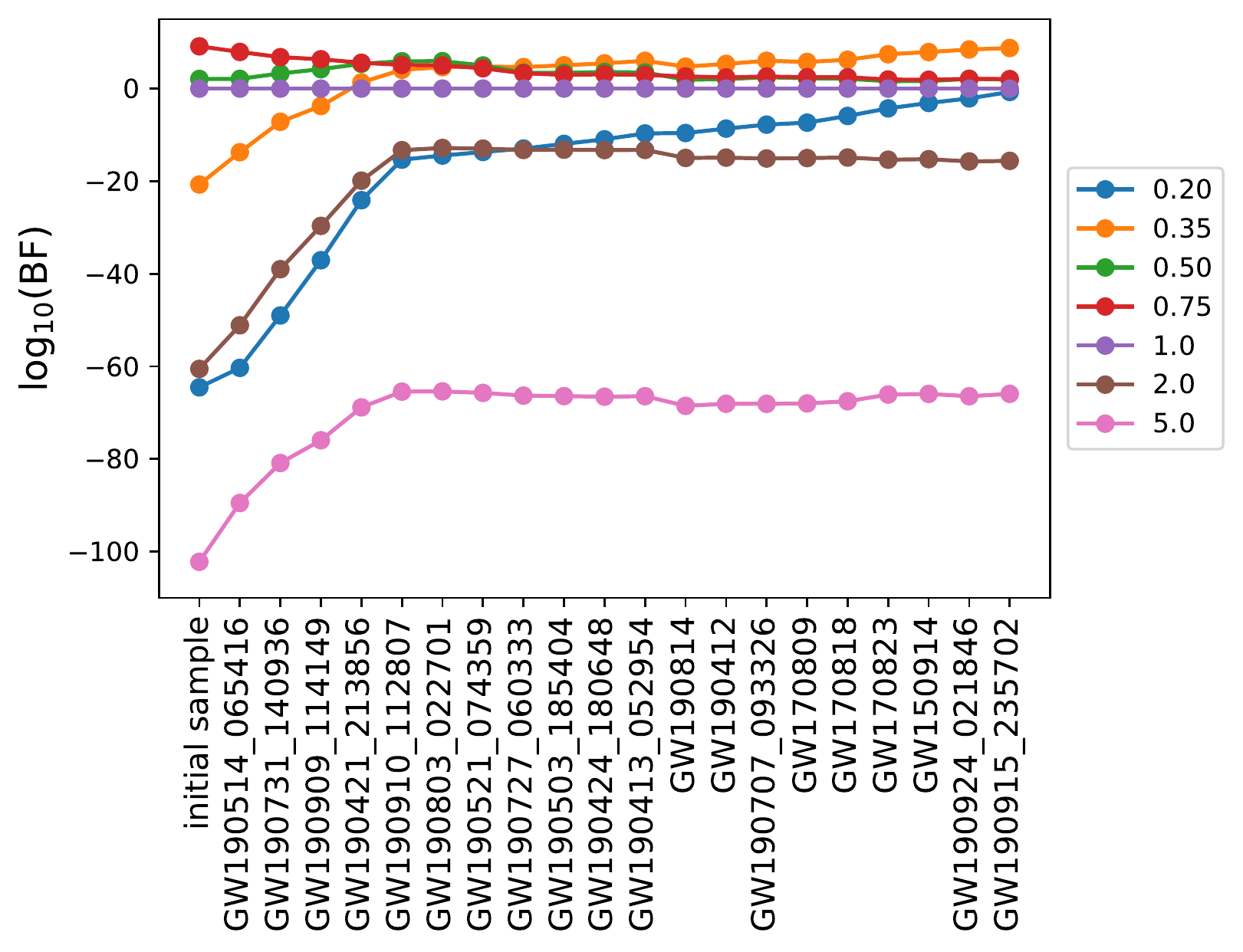}
 \caption{BFs of CE and SMT models with respect to the reference model with $\alpha_\mathrm{CE}=1.0$ and Eddington-limited accretion efficiency as a function of sample size. The initial sample contains $38$ GW events of GWTC-2 and exclude GW170729, GW190413\_134308, GW190519\_153544, GW190521, GW190602\_175927, GW190620\_030421, GW190701\_203306, GW190706\_222641, GW190929\_012149. At each iteration the event with lowest likelihood with respect to the reference model is removed and indicated on the horizontal axis until $20$ events are removed. By definition the BF of the model with $\alpha_\mathrm{CE} = 1.0$ is equal to 1. The data show moderate to strong support for the models with low CE efficiency, $\alpha_\mathrm{CE} < 1.0$. This result is robust because the BFs show a constant behaviour as a function of sample size.
 }
 \label{Fig:BF}
\end{figure}

\subsection{Models comparison}

The real statistical power of model comparison lies in the combined information from all detected events. In Appendix~\ref{app:model-comparison} we explain how to compute the likelihood of observing $N$ independent GW events, described by the physical parameters $\vec{\theta} = (\chieff, \mchirp, q)$, given an astrophysical model described by the set of parameters $\vec{\lambda}$. How well the data are described by one model compared to another is described by the Bayes factor (BF), see Eq.~\eqref{eq:BF}. To compute the BFs, we use as our reference the model of CE and SMT with $\alpha_\mathrm{CE}=1.0$ and Eddington-limited accretion onto the BH. For our model comparisons we remove the GW events discussed in the previous section ({GW170729, GW190413\_134308, GW190519\_153544,} {GW190521, GW190602\_175927, GW190620\_030421}, {GW190701\_203306, GW190706\_222641, GW190929\_012149}) and consider them to not have been formed through the CE and SMT channels in this work. Of course, in the remaining population of events we cannot exclude contamination from other formation channels. A proper analysis would require a model comparison that includes all promising formation channels for BBHs and their branching fractions as model hyperparameters \citep{2020arXiv201110057Z}, this is beyond the scope of this paper as here we are only considering two formation channels.

To estimate which model describes best the events, and how sensitive this result is to the selection of events, we iterate the computation of the BF and remove at each iteration the event with lowest likelihood (with respect to the reference model) until the BF converges to a given value. In Fig.~\ref{Fig:BF} we show the BFs of our reference model to the others as a function of sample size; this converge to a constant value after $5$--$6$ events are removed. The BFs indicate moderate to strong evidence in favour of models with inefficient CE, namely $\alpha_\mathrm{CE} < 1.0$, while excluding the model with lowest $\alpha_\mathrm{CE}=0.2$. If another model is chosen as the reference, then the order of events removed changes, but the end results is the same.

Another question we could ask ourselves is which parameter out of the three (\chieff, \mchirp, $q$) has the most discriminatory power in the BF analysis and, hence, carries most of the information about CE efficiency. To answer this question we repeat the analysis considering each parameter separately in $\vec{\theta}$. We find that the parameter carrying the least information is the mass ratio $q$ which strongly disfavour only the model with $\alpha_\mathrm{CE}=0.2$. Most of the information is contained in \mchirp; considering there is the greatest variation in BFs, for instance, for the model with $\alpha_\mathrm{CE} = 5.0$ the BF is initially disfavoured at $\sim 10^{-12}$ and BF at $\sim 10^{-4}$ for the one with $\alpha_\mathrm{CE} = 2.0$. The only other model disfavoured by the \mchirp dimension is $\alpha_\mathrm{CE}=0.2$ at BF $\sim 10^{-20}$. The \mchirp parameter has the largest discriminating power because is the one affected the most by the variation of $\alpha_\mathrm{CE}$ in the underlying distributions (cf.\ Table~\ref{tab:results} and Table~\ref{tab:relative_fractins}). Larger $\alpha_\mathrm{CE}$ leads to a smaller fraction of systems with $\mchirp > 15 \mathrm{M}_\odot$ which is what most of the LVC data support. In contrast, the \chieff dimension favours models with efficient CE ejection the most: models with $\alpha_\mathrm{CE}=2.0$ at BF $\sim 10^{7}$--$10^{4}$ and $\alpha_\mathrm{CE}=5.0$ at BF $\sim 10^{2.6}$--$10^0$ while it strongly disfavours $\alpha_\mathrm{CE}\in[0.2,0.35]$. This result is a direct consequence of the majority of LVC data supporting small \chieff which in our models are achieved for ultra efficient $\alpha_\mathrm{CE}$. Finally, when we calculate the BFs with $\vec{\theta} = (\chieff, \mchirp)$ we find similar results to the original analysis with $\vec{\theta} = (\chieff, \mchirp, q)$. Moreover, we find that, in contrast to the three-dimensional BF analysis, the two-dimensional one shows rather constant BFs starting from the initial sample. The variation in the first $6$ iterations of the three-dimensional analysis is caused by the $q$ dimension. We should stress, however, that the model with $\alpha_\mathrm{CE} = 0.2$ significantly overpredicts the rate of events. Hence, overall an $\alpha_\mathrm{CE}$ in the range $0.2<\alpha_\mathrm{CE} < 1.0$ is favoured.

The result of our model section analysis needs to be interpret with caution. Here we only explored one parameter of the models which is degenerate with others, for example MT stability and efficiency, BHs birth spin, etc. Moreover we found a non-negligible fraction ($\geq 0.2$) of BBHs originating from other formation channels and, hence, cannot rule out an even greater contamination in the studied population. Finally, other formation channels have been shown to predict BBH observable distributions degenerate with the CE and SMT channels (see references in Sec.~\ref{sec:introduction}). Hence, a proper analysis would require a model comparison of all promising formation scenarios \citep[cf.][]{2020arXiv201110057Z}.


\begin{figure}
    \includegraphics[width=\columnwidth]{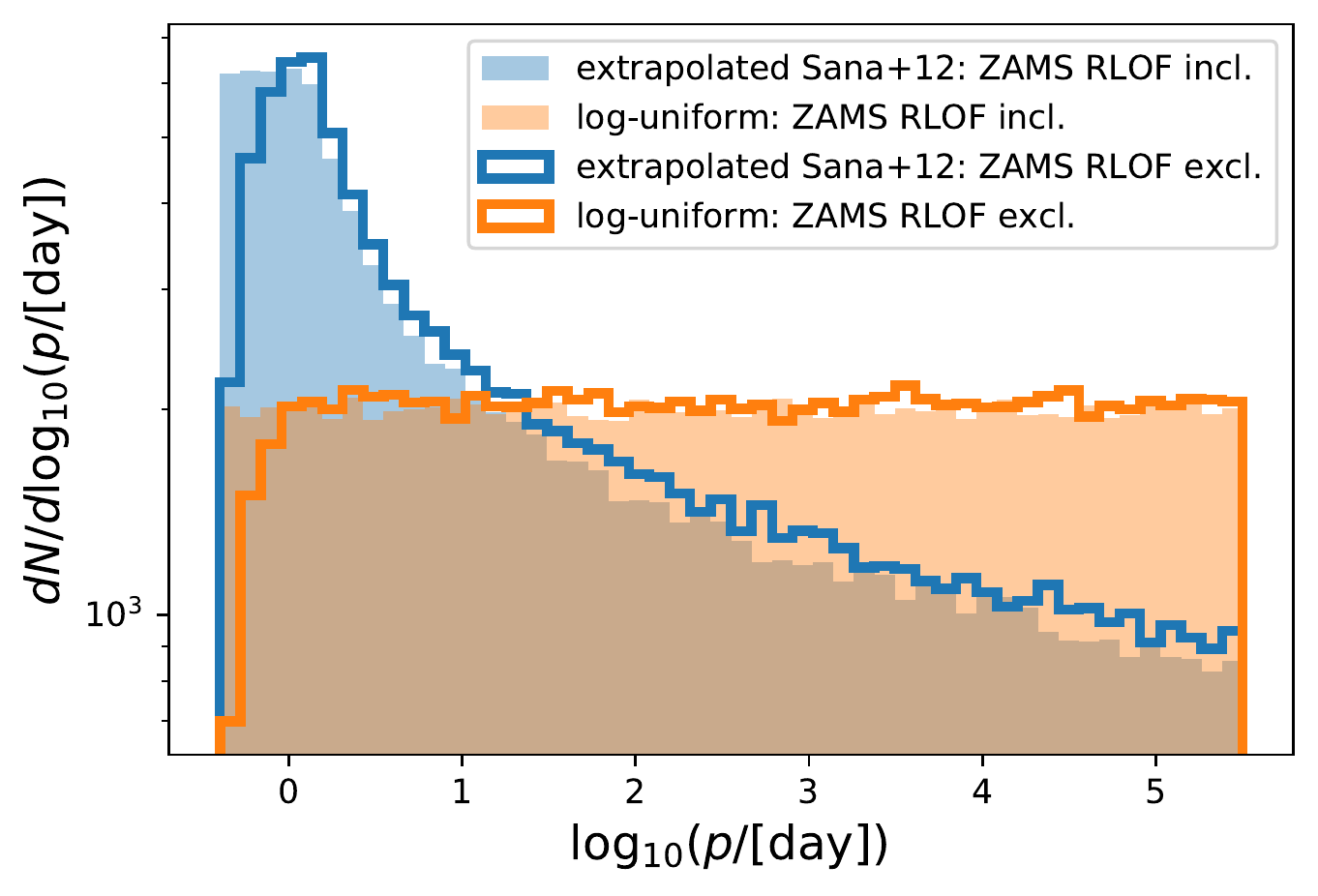}
     \caption{Initial orbital separation of $10^5$ binaries sampled from the extended \citet{2012Sci...337..444S} distribution, cf.\ Eq.~\eqref{eq:Sana}, in blue, and from a log-uniform distribution, in orange. Both distributions sample the range $p \in [0.4,10^{5.5}]$ days and are independent of metallicity. With a solid line, we show the same distributions after removing systems with Roche-lobe overflow at ZAMS for $Z=0.1 \, Z_\odot$.
     }
     \label{fig:psampling}
\end{figure}

\begin{table}
\centering                          
\begin{tabular}{c c c c c}        
\hline\hline                 
  Channel  & orbital  & RLOF  & Rate density  & Detection \\    
  &  dist. & ZAMS & [Gpc$^{-3}$ yr$^{-1}]$ & rate [yr$^{-1}$]  \\    
\hline                       
   CE  & fiducial & incl. & 42.6 & 108 \\
   SMT & fiducial & incl. & 24.6 & 86 \\
   CE  & fiducial & excl. & 55.8 & 142 \\
   SMT & fiducial & excl. & 31.4 & 111 \\
   CE  & log-unif. & incl. & 73.3 & 184 \\
   SMT & log-unif. & incl. & 23.1 & 78 \\
   CE  & log-unif. & excl. & 78.5 & 198 \\
   SMT & log-unif. & excl. & 24.8 & 83 \\
\hline \hline  \\                              
\end{tabular}
\caption{Rate densities at $z=0.01$ and O3 detection rate for SMT and CE models with $\alpha_\mathrm{CE} = 1.0$ and $\eta_\mathrm{acc}=1$ for different initial binary properties. The second column indicates which orbital distribution is used, either log-uniform or fiducial (the extended \citet{2012Sci...337..444S} distribution, see \ Eq.~\eqref{eq:Sana}). The third column indicates if we include or exclude binary systems with Roche-lobe overflow at ZAMS.} 
\label{tab:psampling}
\end{table}

\section{Model uncertainties}\label{sec:modeluncert.}

\subsection{Initial binary properties}\label{sec:initalprop}

Lack of knowledge of the initial binary properties are a source of uncertainty in population synthesis studies. We assume that the primary mass follows the \citet{2001MNRAS.322..231K} initial-mass function (IMF). This is a power law which comes with uncertainties that can affect the rate estimate and to a lesser degree the observed distribution of BBHs \citep[e.g.][]{2015ApJ...814...58D}. Moreover, the IMF may not be universal \citep[e.g.][]{2018Sci...359...69S,2018Sci...361.6506F}. An additional source of uncertainty can be the initial binary mass fraction distribution and birth eccentricities, which we did not investigate here. We expect the impact of these uncertainties to be smaller compared to that of the IMF \citep{2015ApJ...814...58D}.

Another important assumption is the distribution of the birth orbital separation of the binaries. In our model, we assumed it follows an extended \citet{2012Sci...337..444S} power law in log-orbital period $p \in [0.4, 10^{5.5}]$ days, cf.\ Eq.~\eqref{eq:Sana}. Here, we investigate the sensitivity of our results to this assumption. In Fig.~\ref{fig:psampling} we show the histogram of $10^5$ initial orbital separations drawn from the assumed distribution, in blue, compared to a log-uniform distribution, in orange. We find that the rate density of the reference CE model with $\alpha_\mathrm{CE}=1.0$ raises from 42.6~\rdunits to 73.3~\rdunits when assuming log-uniform orbital separation at birth. On the other hand, the rate density of SMT channel remains almost the same. This happens because the log-uniform distribution increases the yield of merging BBHs at large initial orbital periods ($p \gtrsim 10^2$ days), which are the binaries going through the CE evolutionary path, while it does not affect the yield of SMT BBHs as both initial orbital period distributions create roughly the same number of systems at $p \sim 10$ days.

Moreover, the sampled orbital period range can affect our results. In our model we lowered the smallest orbital period, compared to that measured by \citet{2012Sci...337..444S}, to include the portion of the parameter space leading to the chemical homogeneous formation of BBH \citep{2016MNRAS.458.2634M,2016A&A...588A..50M,2020MNRAS.tmp.3003D}. By default, we included systems overfilling their Roche lobe at ZAMS. \citet{2016A&A...588A..50M} found that binaries that are already in contact at ZAMS can potentially survive and lead to the formation of BBHs through chemically homogeneous evolution. However, we only have observations of massive binaries when they are well-past their ZAMS, as in prior evolutionary phases they are still embedded in the formation clouds, undergoing accretion. Accretion onto a pre-MS binary significantly complicates its evolution \citep[e.g.][]{2018arXiv180806488S}, and thus including those binaries in our population models may be problematic. We now investigate how our results change if we exclude these systems. To exclude systems that are initially Roche lobe filling, we adopt the stellar radii fits of \citet{1996MNRAS.281..257T}. These fits are specific for ZAMS and offer more accuracy than the one of \citet{2000MNRAS.315..543H} which are meant to cover the entire stellar evolution and thus sacrifice some accuracy at ZAMS. By removing these binaries, we decrease the number of systems in small orbital periods as shown by the solid lines for both distributions in Fig.~\ref{fig:psampling}. The new way of drawing initial orbital periods is metallicity dependent because, in general, stars have larger radii at larger metallicities \citep{1996MNRAS.281..257T}. In Table~\ref{tab:psampling} we summarise the rates of all these models. As expected, we find for both distributions that excluding RLOF ZAMS increases the rates for both channels. We conclude that this uncertainty can affect our results by a factor of $\lesssim 2$.

\begin{table}
\centering                          
\begin{tabular}{c @{\hspace{0.8\tabcolsep}} c @{\hspace{0.5\tabcolsep}} c @{\hspace{0.3\tabcolsep}} c c c c}        
\hline\hline                 
  &  & Rate& Detection & \multicolumn{2}{c}{$q$} \\    
  channel & $q_\mathrm{crit}$ & density & rate & intrinsic & detected\\
  & & [Gpc$^{-3}$ yr$^{-1}]$ & [yr$^{-1}$] & pop. & pop. \\    
\hline\rule{0pt}{2.6ex}                        
   CE  & F & 42.6  & 108 & $0.79^{+0.17}_{-0.30}$ & $0.82^{+0.14}_{-0.14}$ \\[1mm]
   SMT & F &  24.6 & 86 & $0.72^{+0.20}_{-0.29}$ & $0.74^{+0.12}_{-0.07}$ \\[1mm]
   CE  & C & 50.9 & 124 & $0.78^{+0.18}_{-0.27}$ & $0.81^{+0.15}_{-0.14}$ \\[1mm]
   SMT & C & 118.8 & 354 & $0.66^{+0.22}_{-0.09}$ & $0.69^{+0.14}_{-0.09}$ \\[1mm]
   CE  & B & 24.5 & 28 & $0.79^{+0.18}_{-0.30}$ & $0.84^{+0.13}_{-0.28}$ \\[1mm]
   SMT & B & 31.3 & 212 & $0.94^{+0.05}_{-0.10}$ & $0.94^{+0.05}_{-0.09}$ \\[1mm]
\hline  \hline  \\                                 
\end{tabular}
\caption{Rate densities at $z=0.01$ and O3 detection rate for SMT and CE models with $\alpha_\mathrm{CE} = 1.0$ and $\eta_\mathrm{acc}=1$ for different $q_\mathrm{crit}$ prescriptions. The fiducial model (F) assumes the values presented in Sec. \ref{sec:MT} while the other two options (C) and (B) uses values of \citet{2014A&A...563A..83C} and \citet{2008ApJS..174..223B}, respectively. For comparison, we also report the median BH mass ratio $q$ with its $90\%$ interval.} 

\label{tab:qcrit}
\end{table}

\subsection{Mass-transfer stability}\label{sec:MTstability}

The critical mass ratio $q_\mathrm{crit}$ determines weather the MT is dynamically stable or unstable (cf.\ Sec.~\ref{sec:MT}). We chose our $q_\mathrm{crit}$ values to match the assumptions of our previous work \citep{2020A&A...635A..97B}, which is based on the population synthesis model of \citet{2019MNRAS.490.3740N} obtained using the \compas{} code \citep{2017NatCo...814906S,2018MNRAS.481.4009V}. In contrast to \citet{2019MNRAS.490.3740N}, we implement the same $q_\mathrm{crit}$ fits to the GB and asymptotic AGB \citep{1987ApJ...318..794H} but do not adopt \citet{1997A&A...327..620S} radial response to adiabatic mass loss for evolved stars beyond the HG (this option is not present in the current version of \texttt{COSMIC}). Despite this and other differences in the  model assumptions (such as different CE $\lambda$ fits) which might affect the results, we reached similar conclusions for the detected population of the CE channel with $\alpha_\mathrm{CE} \sim 1.0$. Even though both models converge on similar detected population distributions, the two models have different mass-ratio distributions for the underlying BBH population. We suspect that this discrepancy is caused by the different $q_\mathrm{crit}$ assumption for GB and AGB stars as all our merging BBH systems are evolving through central He burning at onset of CE. In order to verify the source of the difference, a thorough code comparison is needed, which is outside the scope of this project.

To test the sensitivity of our results to this assumption, we run two additional models with different $q_\mathrm{crit}$ choices: model (C) with $q_\mathrm{crit}$ values corresponding to \citet{2014A&A...563A..83C} and model (B) with $q_\mathrm{crit}$ values similar to \citet{2008ApJS..174..223B}. We find model (C) to have similar rate density to our fiducial CE channel but almost five times larger rate density of the SMT channel. On the other hand, the model (B) CE rate density is half of our fiducial model and slightly larger rate densities for the SMT channel. These results are summarised in Table~\ref{tab:qcrit} where we also report for comparison the median mass ratio of the intrinsic and detected BBH populations. Both $q_\mathrm{crit}$ choices do not have a significant impact on the detected observable distributions even though they have different impact on the underlying populations. To determine $q_\mathrm{crit}$, \cosmic{} uses the evolutionary type of the donor (as defined in \citealt{2000MNRAS.315..543H}) rather than the actual structure of its envelope. Recently, \citet{2021A&A...645A..54K} showed that this tends to overpredict the number of systems that evolve through and survive a CE phase. This shows the limitation of parametric population synthesis codes. In fact, $q_\mathrm{crit}$ can be numerically determined by detailed binary evolution simulations, given the profile of the donor star. In the future, next-generation population synthesis tools based on detailed binary and stellar evolution simulations will remove this degree of freedom. 

Finally, in our models we only explore the effects of MT accretion efficiency onto BHs, and did not investigate the effects of MT accretion efficiency between two non-degenerate objects. A recent study \citep{2020arXiv201011220B} showed that, assuming Eddington-limited accretion onto BHs, stars need to accrete more than 30\% of the mass lost by the donor stars during the first MT episode in order to explain O1 and O2 BBHs. MT efficiency between two non-degenerate stars depends strongly on the assumed specific AM of the material that reaches the surface of the mass-gaining star. In turns, this depends on the accretion disk physics and the coupling of the accretion disk to the star's surface. The assumption that the accreted material carries the Keplerian specific AM of the accretor's surface leads to  a very efficient spin up of the mass-gaining star. The accretor then quickly reaches critical rotation and halts further accretion, leading to a highly inefficient mass transfer \citep[a few to a few tens percent; e.g.][]{2013ApJ...764..166D,2020A&A...638A..39L}. On the other hand, if one assumes that the AM is dissipated efficiently before it reached the accretor's surface, and that that the material that is added onto the star has similar specific AM to that of its own surface layers, then MT can be significantly more efficient.

\subsection{Effect of angular momentum transport \& accretion efficiency on black hole spin}\label{sec:qcrit}

Our models assume efficient AM transport \citep{1999A&A...349..189S,2002A&A...381..923S} which leads to the formation of non-spinning first-born BHs \citep{2015ApJ...800...17F, 2018A&A...616A..28Q}. Although the Tayler--Spruit dynamo model helps to reproduce the flat rotation profile of our Sun \citep{2014ApJ...796...17F,2014ApJ...788...93C}, as well as, neutron star and white dwarf spins \citep{2005ApJ...626..350H,2008A&A...481L..87S}, it cannot reproduce the asteroseismic constrains for subgiants and red giants \citep{2018A&A...616A..24G}. This would require an even higher efficiency in AM transport. Alternatively, a model with inefficient AM transport predicts highly spinning BHs \citep[e.g.][]{2019MNRAS.482.2991A,2020A&A...636A.104B}, which do not match current GW observations. If the second MT episode is stable, then the first-born BH can be spun up by accretion \citep{1974ApJ...191..507T}. If the MT accretion onto BHs is Eddignton-limited, the BH accretes a negligible amount of matter leading to $a_1 \simeq 0$. On the other hand, a super Eddington-limited MT accretion will result in larger spins. The extreme case of this would be highly super-Eddington accretion efficiency where the spin of the first-born BH can even approach to  $a_1 \simeq 1$, but in this case the contribution to the merging BBH population of the SMT channel almost vanishes (Table~\ref{tab:results}).

The spin of the second-born BH is determined by the combined effects of stellar winds and tidal interactions during the BH--He-star binary evolution which we model in detail. During this evolutionary phase, the AM transport does not play an important role as the He-star is compact and will not expand during its final evolution \citep{2020A&A...635A..97B}. The strength of the tidal interaction is primarily determined by the orbital separation during the BH--He-star evolutionary phase, see Eq.~\eqref{eq:tides}. In our SMT models, the orbital separation is determined by the accretion efficiency. Models where super-Eddington accretion is allowed will result in larger orbits than our reference model, decreasing further the small effect of tides on this evolutionary path. 

\subsection{Common-envelope prescription}

In our CE models, the post-CE orbital separation, $A_\mathrm{postCE}$, is linearly dependent on the CE parameterisation uncertainties as, approximately, $\delta A_\mathrm{postCE} / A_\mathrm{preCE} \propto \delta \alpha_\mathrm{CE} \, \lambda$. Even though we did not explore different $\lambda$ fits in our models, our parameter study of $\alpha_\mathrm{CE} \in [0.2,5.0]$ covers an uncertainty on $ \delta A_\mathrm{postCE} / A_\mathrm{preCE} \propto 5.0/0.2 \simeq 25$. Recently, \citet{2021A&A...645A..54K} showed that $\lambda$ fits similar to the one we used could underestimate the envelope binding energies of massive radiative-envelope giants, leading to an overestimation of the systems surviving CE. An additional free parameter in the calculation of $\lambda$ which complicates a detailed treatment of CE is the assumed boundary down to which the envelope will be ejected. Unfortunately detailed stellar models cannot robustly predict this \citep[e.g.][]{2001A&A...369..170T} and hydrodynamic simulations of the CE phase are necessary. For example, \citet{2019ApJ...883L..45F} showed that for progenitors of binary neutron stars, a non-negligible fraction of hydrogen rich material will remain bound to the core after the successful ejection of the CE, that would in turn imply a relatively efficient ejection of the CE. 

\subsection{Other uncertainties}

Our model may be limited by other uncertainties we did not explore which can alter the merger rate and, to a lesser degree, the predicted BBH property distributions. Uncertainties in the (i) physics of the supernova
explosions, such as the kicks strength, can influence rates and affect the parameter distribution of BBH mergers \citep[e.g.][]{2013ApJ...779...72D}. When connecting the population synthesis code to our detailed \mesa{} simulations we (ii) assumed the BH--He-star systems post second MT to be at ZAHeMS. This is not always the case as the binaries are evolving through central He burning at onset of the MT. This leads us to overestimate the lifetime of these He-stars. This overestimation is negligible as the binary only spends a few hundred thousand years in this state compared to its overall life of a few million years and much longer BBH inspiral. This overestimation might lead to less massive second-born BHs and smaller spins as winds act for a larger time window. However, we expect that the fraction of stars entering the MT on advanced He-burning phase is higher at low metallicities, as low-metallicity stars tend to expand later in their lives \citep{2010ApJ...725.1984L}. At the same time, stellar winds in these stars are weaker due to the low metallicity, so the overall effect on the population of BBHs is expected to be limited. The uncertainty of the (iii) metallicity dependence of stellar winds for massive stars is another important ingredient of population synthesis studies which affects \mchirp distributions and the rates \citep[cf.][]{2018MNRAS.477.4685B}. The detection rate and density calculation is also affected by uncertainties in the (iv) redshift and metallicity dependent SFR \citep{2019MNRAS.482.5012C, 2019MNRAS.490.3740N,2020MNRAS.493L...6T}. A SFR favouring higher formation metallicities than the one assumed here would skew our results in favour of smaller \chieff and lower rates as low metallicity systems are responsible for high \chieff and the short merger timescales. 

\section{Conclusions}\label{sec:conclusions}

Mass-transfer physics is one of the most uncertain physical processes determining the observable properties of field binary black holes. The critical mass ratio $q_\mathrm{crit}$ determines the fraction of the parameter space going trough SMT and CE phases. Mass-accretion efficiency onto compact objects determines how efficiently binaries going through SMT will shrink, while CE efficiency $\alpha_\mathrm{CE}$ determines post-CE orbital separations. In this work we investigated how the detected joint distributions of \mchirp, \chieff and $q$ of BBH formed through the CEe and SMT formation channels change given the uncertainties on these input physics. We investigated this by combining rapid binary population synthesis studies with detailed stellar and binary simulations. Rapid population synthesis studies allow us to obtain different BH--He-star populations for different input physics, while detailed simulations which take into account differential stellar rotation, tidal interaction, stellar winds and the evolution of the He-star stellar structure allow us to accurately determine the distributions of BBH observables. We then convolved the synthetic BBH population with the redshift- and metallicity-dependent star-formation rate, as well as selection effects from a 3-detector network to build a model capable of describing LIGO--Virgo detections. Our main findings are:
\begin{itemize}
    
    \item We calculated the O3 detected joint distributions of \chieff, \mchirp and $q$ for CE and SMT channels. Assuming efficient AM transfer inside stars and Eddington-limited accretion efficiency, both channels lead to similar rate densities in the local Universe. We find that the CE channel is the only evolutionary path leading to non-zero \chieff in the detected population as SMT channel cannot shrink the orbits enough for efficient tidal spin-up to take place.

    \item Inefficient CE (small $\alpha_\mathrm{CE}$ values) leads to smaller orbital separation post CE. On average, these models lead to more systems being tidally spun up. However, the majority of these systems are not detected by current GW detectors because most of these systems are formed in low metallicity environments (otherwise stellar winds widen the binaries) and merge quickly at a redshift close to their formation ($z \sim 2$ where the SFR peaks), far outside current GW detector horizons ($z \lesssim 1$).
    
    \item Highly super-Eddington accretion efficiency onto compact objects reduces the rate densities of CE by $\sim 10\%$, while it reduces the contribution of SMT channel by two orders of magnitude, making the contribution of this channel almost negligible compared to the CE channel.
    
    \item The GW events {GW170729, GW190413\_134308,} {GW190519\_153544, GW190521, GW190602\_175927,} {GW190620\_030421, GW190701\_203306,} GW190706\_222641 and GW190929\_012149 of  GWTC-2 are not well explained by our models of isolated binary evolution through either CE or SMT. If we assume that these systems originate from other formation channels then we can put a lower bound on the detected branching fraction from other formation channels: $9/47\simeq 0.2$.
    
    \item We conducted a model comparison given the events of GWTC-2 consistent with our CE and SMT models to determine which CE efficiency is best supported by the data. The GW events show moderate to strong evidence in favour of the models with inefficient CE, $0.2 < \alpha_\mathrm{CE} < 1.0$. We find this result to be robust considering different selections of events in our calculations. This analysis did not include rate estimates which the $\alpha_\mathrm{CE}=0.2$ model significantly overpredicts.
    
\end{itemize}
We conclude that future works aiming to properly infer model parameters through model comparison will need to consider correlation between parameters as well as contamination from other formation channels in order to properly determine model parameters.

\begin{acknowledgements}
     We would like to thank Simon Stevenson and Katerina Chatziioannou for useful discussions. 
     This work was supported by the Swiss National Science Foundation Professorship grant (project number PP00P2 176868). This project has received funding from the European Union's Horizon 2020 research and innovation program under the Marie Sklodowska-Curie RISE action, grant agreement No 691164 (ASTROSTAT). MZ is supported by NASA through the NASA Hubble Fellowship grant HST-HF2-51474.001-A awarded by the Space Telescope Science Institute, which is operated by the Association of Universities for Research in Astronomy, Incorporated, under NASA contract NAS5-26555. CPLB is supported by the CIERA Board of Visitors Research Professorship. PM is supported by the FWO junior postdoctoral fellowship No.\ 12ZY520N. JJA and SC are supported by CIERA and AD, JGSP, and KAR are supported by the Gordon and Betty Moore Foundation through grant GBMF8477. KK received funding from the {\it European Research Council} under the European Union's {\it Seventh Framework Programme} (FP/2007-2013) / {\it ERC} Grant Agreement n.~617001. YQ acknowledges funding from the Swiss National Science Foundation under grant P2GEP2{\_}188242. The computations were performed in part at the University of Geneva on the Baobab and Lesta computer clusters and at Northwestern University on the Trident computer cluster (the latter funded by the GBMF8477 grant). All figures were made with the free Python modules Matplotlib \citep{Hunter:2007} and pygtc \citep{Bocquet2016}. This research made use of Astropy,\footnote{\href{http://www.astropy.org}{www.astropy.org}} a community-developed core Python package for Astronomy \citep{astropy:2013,astropy:2018}. 
\end{acknowledgements}

%
%

\bibliography{aanda}

\appendix 

\section{Initial binary distributions}\label{app:initial_properties}

The parameters describing the initial conditions of a binary system are primary  and secondary masses, $m_1$ and $m_2$, orbital period $p$, eccentricity $e$ and metallicity $Z$ of the stars at ZAMS. We assumed that the primary masses follow the initial mass function (IMF) of \citet{2001MNRAS.322..231K} which spans the mass range $0.01 \, \mathrm{M}_\odot \leq m_1 \leq 150 \, \mathrm{M}_\odot$. The upper limit is an extrapolation of the \citet{2001MNRAS.322..231K} IMF which is measured only up to $50 \, \mathrm{M}_\odot$. In our model, the lower limit represents the smallest theoretical mass for a star to support nuclear fusion \citep[cf.][]{1963ApJ...137.1121K} while the arbitrary maximum stellar mass exclude BH formation above the upper mass-gap \citep[e.g.][]{2002luml.conf..369H}. The mass distribution of the less massive secondary star is given by $m_2 = m_1 q$ where the initial mass ratio $q$ is drawn from a flat distribution \citep{2012Sci...337..444S} in the range $q \in [0,1]$. Furthermore, we adopt a binary fraction of $f_\mathrm{bin}=0.7$ \citep{2012Sci...337..444S} and assume that at birth the distribution of log-orbital periods follow \citet{2012Sci...337..444S} power law with coefficient $\pi = - 0.55$ in the range $p \in [10^{0.15},10^{5.5}]$ days and extrapolate down to the range $p \in [0.4,10^{0.15}]$ days assuming a log-flat distribution (as  the power low is not defined for $p < 10^{0.15}$ days), that is
\begin{equation}
    f(p) = \begin{cases} 
      C \times 0.15^{-0.55}  &  0.4 \leq p / \mathrm{days} < 10^{0.15} \\
      C \log_{10}(p/ \mathrm{days})^{-0.55}  &  10^{0.15} \leq p / \mathrm{days} \leq 10^{5.5}
   \end{cases},
   \label{eq:Sana}
\end{equation}
where the normalisation constant $C$ is determined by the condition $\int_{p_\mathrm{min}}^{p_\mathrm{max}} f(p) \, \mathrm{d}p = 1$.
The lower limit of $0.4$ days will ensure that we probe all the available parameter space, see for example, the detailed simulations of \citet{2020MNRAS.tmp.3003D} where they find binaries with initial orbital periods as small as $0.4$ days forming BBHs. Finally, we assume that all binaries are born with circular orbits, namely with zero $e$. We assume that all these distributions are both independent of each other, as well as, independent of metallicity which in reality might not be the case \citep{2017ApJS..230...15M}. This oversimplification means that we also include systems overfilling the L1 and L2 Roche-lobe surfaces at ZAMS. In Sec.~\ref{sec:initalprop} we discuss how our BBH rate estimates are affected by omitting these systems or by assuming a log-uniform distribution. The omission of this portion of the parameter space does not lead to a qualitative difference in the predicted observable distributions.

In order to optimise the absolute number of binaries becoming BBHs per number of simulated binaries, we restrict the primary mass to the range $5 \, \mathrm{M}_\odot \leq m_1 \leq 150 \, \mathrm{M}_\odot$. For the metallicity, we divide uniformly the log-metallicity range $Z \in [0.0001, 0.0309]$ in 30 bins. The largest metallicity bin was chosen to have a centre at $1.5 \, Z_\sun$ where we adopt the solar reference $Z_\odot = 0.017$ \citep{1996ASPC...99..117G}. This metallicity range is where the stellar model fits of \citet{2000MNRAS.315..543H} we use are defined. We evolve 5 million binaries per metallicity bin $\Delta Z$ for a total simulated mass per $\Delta Z$ of $M_{\mathrm{sim}, \, \Delta Z} \simeq 1.05 \times 10^{8} \, \mathrm{M}_\odot$. Since we restricted the primary mass, we only model a fraction of the underlying stellar population. Hence, we need to re-normalise the simulated stellar mass $M_{\mathrm{sim}, \, \Delta Z}$ to obtain the total stellar population; the normalisation constant for our choice of initial binary properties is $f_\mathrm{corr}^{-1}=4.72$, see Appendix~A in \citet{2020A&A...635A..97B} for a derivation of this quantity.

\section{COSMIC population synthesis model}\label{app:COSMIC}

\subsection{Single stellar models}

To generate our BBH population models, we used \cosmic{} version \texttt{v3.3.0}. Stellar evolution in \cosmic{} \citep{2020ApJ...898...71B} is based on the analytical fits of \citet{2000MNRAS.315..543H,2002MNRAS.329..897H} to the single stellar models of \citet{1998MNRAS.298..525P}. For O and B stars we adopt mass loss through stellar winds according to the prescription of \citet{2001A&A...369..574V}, which covers separately the temperature ranges $12,500 \,  \mathrm{K} < T_\mathrm{eff} < 22,500 \, \mathrm{K}$ and $27,500 \, \mathrm{K} < T_\mathrm{eff} < 50,000 \mathrm{K}$. Around $T \simeq 25,000 \, \mathrm{K}$ there is a bi-stability jump that leads to a mass-loss increase of a factor of about five. This jump is due to the recombination of the Fe IV to the Fe III ion in the lower part of the wind \citep{1999A&A...350..181V}. For Wolf--Rayet stars mass-loss rate we adopt stellar winds as in \citet{2011A&A...530A.115B} who assume \citet{1995A&A...299..151H} reduced by a factor of $10$ to correct for clumping and use a metallicity scaling of $(Z/Z_\odot)^{0.85}$ \citep{2001A&A...369..574V}. For all these wind prescriptions we adopt the solar reference $Z_\odot = 0.017$ \citep{1996ASPC...99..117G}.

During the post-carbon burning phase of massive stars, photons produced in the core are energetic enough to produce electron--positron pairs which softens the equation of state, diminishing the pressure support of the core \citep{2007Natur.450..390W}. This causes the core to rapidly contract and the temperature to increase, allowing for explosive oxygen burning \citep[e.g.][]{2015ASSL..412..199W}. For He-core masses in the range $\sim [30,64] \, \mathrm{M}_\odot$ the released energy is insufficient to completely disrupt the star. This create a series of energetic pulses which eject material from the star before it collapses into a BH. This is the PPI \citep[][]{2016MNRAS.457..351Y,2017ApJ...836..244W,2019ApJ...882...36M}. If the He-core mass is in the range $\sim [64,133] \, \mathrm{M}_\odot$ the released energy is enough to reverse the collapse, unbinding and destroying the star. This event is a PISN  \citep[][]{1964ApJS....9..201F,1967ApJ...148..803R,1967PhRvL..18..379B}. Similar to \citet{2019ApJ...882..121S}, we adopt the fit to the grid of simulations from \citet[][see Table 1]{2019ApJ...882...36M}, which demonstrate a turnover in the relation between pre-supernova He-core mass and final mass. We use the 9th-order polynomial fit of \citet{2020ApJ...898...71B}, cf. their Eq.~(4), to map the pre SN stellar mass in the range $31.99 \leq M_\mathrm{preSN}/\mathrm{M}_\odot \leq 61.10$ to the baryonic mass collapsing to form the BH.

\subsection{Mass-transfer stability and common envelope}\label{sec:MT}

The stability of Roche-lobe overflow mass transfer is determined by the rate at which the Roche-lobe radius is changing as a result of mass-transfer $\mathrm{d} \log(R_\mathrm{L})/\mathrm{d} \log(m)$ to the response of the radius of a star as its mass is changing $\mathrm{d} \log(R_*)/\mathrm{d} \log(m)$. We use the approximation of \citet{1983ApJ...268..368E} for the Roche-lobe radius while we approximate the radial response of the star depending on its stellar type. We adopt the values assumed in \citet{2019MNRAS.490.3740N,2020A&A...635A..97B}. The stability of the mass transfer can then be determined by solving this equation with respect to the critical mass ratio, defined as $q_\mathrm{crit} = m_\mathrm{donor}/m_\mathrm{accretor}$. For MS stars we use $\mathrm{d} \log(R_*)/\mathrm{d} \log(m) = 2.0$ which correspond to $q_\mathrm{crit} \simeq 1.72$ while for HG stars $\mathrm{d} \log(R_*)/\mathrm{d} \log(m) = 6.5$ which correspond to $q_\mathrm{crit} \simeq 3.83$ \citep{2015ApJ...812...40G}. For stars on the GB and AGB we use fits from \citet{1987ApJ...318..794H}. For stripped stars we adopt $q_\mathrm{crit}$ as in \citet{2014A&A...563A..83C}. Different choices of $q_\mathrm{crit}$, especially for GB and AGB stars have an impact on the parameter space that leads to the formation of BBHs, hence on the merger rate, see Sec.~\ref{sec:qcrit} for a discussion of this uncertainty. 

If the mass transfer is stable the companion star accretes a fraction of mass lost by the donor star. Any mass that is not accreted leaves the system instantaneously, taking away the specific AM of the accretor \citep{2002MNRAS.329..897H}. For our fiducial models we assume that the accretion of degenerate objects is Eddington-limited, this results in a highly non-conservative mass-transfer phase where degenerate objects accrete  negligible amount of  mass, hence the first-born BH will not spin up because of accretion during the second mass-transfer phase \citep{1974ApJ...191..507T}. For the Eddington-limited accretion efficiency onto a compact object, \cosmic{} uses the definition of \citet{2002MNRAS.329..897H},
\begin{equation}
    \dot{M}_\mathrm{edd} = 2.08 \times 10^{-3} (1+X)^{-1} (R_\mathrm{acc}/\mathrm{R}_\odot) \, \, \, \mathrm{M}_\odot \, \mathrm{yr}^{-1} \equiv 1 = \eta_\mathrm{acc} \, ,
\end{equation}
where $X$ is the hydrogen mass fraction of the donor and $R_\mathrm{acc}$ is the accretion radius of the compact object which for a BH is chosen to be the Schwarzschild radius. In this work we define $\eta_\mathrm{acc}$ to be the MT accretion efficiency limit in units of the Eddington-limit. To investigate how our results depend on this limit, we explore different MT efficiency limits: $ \eta_\mathrm{acc} \in [1, 10^3, 10^5, 10^9]$.

If the mass transfer is unstable, the donor star will expand to form a CE of gas around the binary which can be expelled by the injection of orbital energy from the binary \citep{1976IAUS...73...75P}. This is a complex phase and we parameterise it with the $\alpha_\mathrm{CE}$--$\lambda$ formalism \citep[see, e.g.][for a review]{2013A&ARv..21...59I}. In this parameterisation $\alpha_\mathrm{CE}$ determines the efficiency factor for injecting orbital energy into the envelope, while $\lambda$ characterises the binding energy of the envelope to its stellar core which depends on
the structure of the donor's envelope. Assuming that the pre-CE orbital energy is much smaller than the post-CE orbital energy, the initial and final orbital period of the CE event, $A_\mathrm{preCE}$ and $A_\mathrm{postCE}$, are related by the following expression \citep{2013A&ARv..21...59I}
\begin{equation}
    A_\mathrm{postCE} \simeq \frac{1}{2}  \alpha_\mathrm{CE} \,  \lambda  \frac{m_\mathrm{1} \, m_\mathrm{2,postCE}}{m_\mathrm{2,preCE} \, m_\mathrm{2,env}} \, \hat{r}_L \, A_\mathrm{preCE},
    \label{eq:CE}
\end{equation}
where $\hat{r}_L$ is the dimensionless Roche-lobe radius, $m_\mathrm{1}$ is the mass of the accretor, $m_\mathrm{2,preCE}$ and $m_\mathrm{2,postCE}$ are the donor star masses before and after the CE event and $m_\mathrm{2,env}$ the envelope mass. 
Previous studies of post-CE binaries have shown that the efficiency could be as low as $\alpha_\mathrm{CE} = 0.2$ \citep{2010A&A...520A..86Z, 2013A&A...557A..87T,2014A&A...566A..86C} while other using detailed modelling of the CE phase for binary neutron star progenitors \citep{2019ApJ...883L..45F}, suggests CE efficiencies as high as $\alpha_\mathrm{CE} = 5.0$ \citep[see also][]{2019MNRAS.482.2234G}. Approximately, uncertainties on $\alpha_\mathrm{CE}$ or on $\lambda$ linearly scale to uncertainties on the orbital separation post CE as $\delta A_\mathrm{postCE} / A_\mathrm{preCE} \propto  \delta \alpha_\mathrm{CE} \times \lambda + \alpha_\mathrm{CE} \times \delta \lambda$.
For $\lambda$ we adopt the fits as in \citet{2014A&A...563A..83C}.

Within the CE channel we distinguish and adopt a pessimistic scenario, in which unstable mass transfer from a donor star without a well-developed core-envelope structure, namely when a star finds itself in the HG, is always assumed to lead to a merger \citep{2007ApJ...662..504B}. An optimistic scenario which include these systems would result in an overestimation of the observed local BBHs merger rate density \citep{2019MNRAS.490.3740N,2020A&A...636A.104B}.

\section{MESA detailed BH--He-star models}\label{app:MESA}

In our previous works \citep{2018A&A...616A..28Q,2020A&A...635A..97B} we explored the evolution of tight BH--He-star systems computing grids of detailed binary evolution models using \mesa{} \citep{2011ApJS..192....3P,2013ApJS..208....4P, 2015ApJS..220...15P,2018ApJS..234...34P,2019ApJS..243...10P}. This is the late-end phase of the binary evolution of BBHs formed through the CE and SMT channels. Here we iterate on this work by adapting those models to the newer \mesa{} version \texttt{r11701}, computing an even larger grid and modifying our stellar models to match the physical stellar assumptions made in the work of \citet{2020MNRAS.tmp.3003D}, which expand on the work of \citet{2016A&A...588A..50M} on chemical homogeneous evolution.

\subsection{Single stellar physics}

All mixing processes are treated as diffusive processes.
Convection is modelled using the Ledoux criterion, adopting a mixing-length parameter of $\alpha_\mathrm{MLT} = 1.5$ \citep{1958ZA.....46..108B}.
Semiconvection is modelled according to \citet{1983A&A...126..207L} with an efficiency parameter $\alpha_\mathrm{SEM} = 1.0$ \citep{1991A&A...252..669L}.
We also take into account exponential core-overshooting and thermohaline mixing with $\alpha_\mathrm{TH} = 1.0$ \citep[cf.][]{2009A&A...499..279C}. Opacities are computed using CO-enhanced opacity tables from the OPAL project \citep{1996ApJ...464..943I}, computed using solar-scaled abundances based on \citet{1996ASPC...99..117G}.
As we are not interested in following the nucleosynthesis in detail, we use the simple networks provided with \mesa{} \texttt{basic.net} which \mesa{} automatically extend to \texttt{co\_burn.net} to resolve carbon burning. Our models are stopped at C depletion. Finally, PPI and PISN are treated as in the single stellar models of \texttt{COSMIC}. We assume that the mass loss through PPI of any stellar model is lost from the surface of the star and only consider the AM of the remaining layers in the collapse of the star.

Rotational mixing and AM transport are also treated as diffusion processes \citep{2000ApJ...528..368H,2005ApJ...626..350H}, which involve the effects of Eddington--Sweet circulation, secular and dynamical shear mixing and the Goldreich--Schubert--Fricke instability with an efficiency parameter of $f_c = 1/30$ \citep{1992A&A...253..173C,2000ApJ...528..368H}. We include the effect of magnetic fields on the transport of AM assuming an efficient AM transport mechanism: the Tayler--Spruit dynamo \citep{1999A&A...349..189S,2002A&A...381..923S}. An efficient AM transport allows us to assume that all He-stars emerging from the CE or SMT phases are initially not rotating. This is because any initial or acquired rotation during the evolution of the secondary is erased by mass transfer and wind mass loss by the time it becomes a He-star. Assuming instead that the He-star is initially synchronized with the orbit, right after the formation of the BH--He-star binary, has been shown to have negligible effects in the final properties of the resulting BBH system \citep{2018A&A...616A..28Q}.

We implement the same WR stars wind prescription as in \cosmic{} where we also include the enhancement of winds through rotation as in \citet{2000ApJ...544.1016H}. When the rotation rate exceeds a given threshold, $\Omega / \Omega_\mathrm{crit}> 0.98$, we implicitly compute the mass-loss rate required for the rotation rate to remain just below this value \citep[][]{2015ApJS..220...15P}. For a star with mass M and radius R, $\Omega_\mathrm{crit} \equiv \left[(1-\Gamma)GM/R\right]^{1/2}$ where $\Gamma = \kappa L / (4\pi c G M)$ is the Eddington factor and $\kappa$ is the true flux-mean opacity coefficient \citep{1997ASPC..120...83L}.

\subsection{Binary stellar physics}

We use \mesa{} single star module to first create a He-star with the desired mass and with abundances $Y=0.98$ and $Z=0.02$. We then load the model in the \mesa{} binary module, relax the metallicity to the desired value and allow the star to relax until it reaches ZAHeMS. We define ZAHeMS to be the moment when the central luminosity becomes larger than $\sim99\%$ of the surface luminosity. To facilitate the relaxation of the star to ZAHeMS we adopt \texttt{MLT++} treatment of convection \citep{2013ApJS..208....4P} which reduces the superadiabaticity in some radiation-dominated convective regions. Once the star reaches ZAHeMS we turn off \texttt{MLT++}. At ZAHeMS we check if the He-star overfill its Roche lobe, if this is the case we stop the evolution and assume the system to be non-physical. Whenever one component in the system attempts to overflow its Roche lobe during the binary evolution, we implicitly compute the mass-transfer rate using the Kolb scheme \citep{1990A&A...236..385K}. If the mass transfer exceed 10 M$_\odot\,\mathrm{yr}^{-1}$ we stop the run and assume that the binary will merge. Furthermore, we use the prescription of \citet{2020A&A...642A.174M} to check that the He-star does not overflow the L2 Roche volume of the binary. If this is the case, either at ZAHeMS or during the evolution of the binary, we stop the run and assume that it will lead to a merger.

Tidal forces are responsible for synchronising the spin of the He-star with the orbit \citep{1977A&A....57..383Z,1981A&A....99..126H}.
Tidal effects are implemented as in \citet{2019ApJ...870L..18Q} for the case of stars with radiative envelope. This is a variation of the standard tidal prescription of \mesa{} \citep{2015ApJS..220...15P} which synchronize the whole star. Here tides only operate on the radiative regions. This slight variation has a negligible impact on our results because the Tayler--Spruit dynamo guarantees strong coupling between the star's layers. We assume that the orbits after the second MT phase are circular and the system remains circular during binary evolution. The strength of the interaction depends on the ratio of the stellar radius $R$ to the orbital separation $A$, the He-star mass $M$, the binary mass ratio $q$ and the moment of inertia $I$. The timescale for synchronization is given by \citep{2002MNRAS.329..897H}
\begin{equation}
    \frac{1}{T_\mathrm{syn}} = 3 E_2 (1+q)^{5/6} \frac{q^2}{r_\mathrm{g}^2}\left( \frac{GM}{R^3} \right)^{1/2} \left( \frac{R}{A} \right)^{17/2} , 
    \label{eq:tides}
\end{equation}
where $r_\mathrm{g} = [I/(MR^2)]^{1/2}$ is the dimensionless gyration radius of the He-star and $E_2$ is the second order tidal coefficient. As in our previous work, we use the fitting formula of $E_2
$ for He-star from \citet{2018A&A...616A..28Q}. 

\section{Core collapse}\label{app:corecollapse}

BHs are formed during the core collapse of massive stars and, in some cases, their formation is associated with a supernova explosion. As in \citet{2020A&A...635A..97B}, we use \citet{2012ApJ...749...91F} delayed supernova prescription to model how much baryonic remnant mass is left behind after the core collapse. This model avoids an enforced mass gap between neutron stars and BHs which are not consistent with current microlensing observations \citep{2020A&A...636A..20W} or with GW190814 \citep{2020ApJ...896L..44A}. During the collapse of the star, asymmetric ejection of matter \citep{1994A&A...290..496J,2013RvMP...85..245B} or asymmetric emission of neutrinos \citep{1993A&AT....3..287B,2005ApJ...632..531S} can provide a momentum kick to the newly formed BHs. We assume that the birth kicks of BHs follow a Maxwellian distribution with $\sigma = 265~\mathrm{km\,s^{-1}}$ \citep{2005MNRAS.360..974H} rescaled by $1-f_\mathrm{fb}$ where $f_\mathrm{fb}$ is the fallback mass fraction \citep{2012ApJ...749...91F}. For massive stars with carbon--oxygen core masses grater than $11~\mathrm{M}_\odot$, \citet{2012ApJ...749...91F} prescription assumes a direct collapse, that is $f_\mathrm{fb}=1$. This implies that in our model all heavy BHs receive no natal kicks. The kicks impacted on the lighter BHs can tilt the orbit, which may generate a negative \chieff, add eccentricity to the orbit or disrupt the binary. In practise systems with negative \chieff are not statistically relevant in our models as the kicks are not strong enough to flip the orbits by more than $90^\circ$. Recently, \citet{2020arXiv201109570C} showed that a $\sigma\simeq 1,000$ km/s is required in order to explain negative \chieff in GWTC-2 data with CE BBHs. We take into account all these orbital changes, as well as, orbital changes due to neutrinos symmetric mass loss, following the analytical calculations of \citet{1996ApJ...471..352K}.

We estimate the spin of the resulting second-born BH following
the framework presented in \citet{2020A&A...635A..97B}. Here, we adopt a different treatment of neutrino mass loss motivated by \citet{2020ApJ...899L...1Z} which prescribes that a collapsing proto-neutron star can lose up to $0.5 \, \mathrm{M}_\odot$ through neutrino emission. In order to calculate the final mass and spin of the BH resulting from the collapse, we need to consistently follow its accretion history soon after it is formed.
We assume that the innermost $3\, \mathrm{M}_\odot$ form a proto-neutron star, which collapse to form a BH of $2.5 \, \mathrm{M}_\odot$, while $0.5\, \mathrm{M}_\odot$ are converted into neutrinos and leave the systems carrying away a fraction ($0.5/3$) of the proto-neutron star AM.
We consider a collapsing star to be a collection of shells with mass $m_\mathrm{shell}$ and angular frequency $\Omega_\mathrm{shell}$ that falls one by one onto the central BH. Once a shell reaches the BH’s event horizon, it is accreted by it.
The amount of specific AM of the in-falling material, $j(r,\theta) = \Omega_\mathrm{shell}(r) r^2 \sin(\theta)$ where $\theta$ is the polar angle, determines the properties of the accretion flow.
Low AM material collapses directly onto the BH transferring its entire mass and AM to the hole, while material with enough AM can create a disk around it. The mass $M_\mathrm{BH}$ and spin $a$ of the accreting BH determine the threshold for disk formation and is given by the specific AM at the innermost stable circular orbit \citep[ISCO;][]{1972ApJ...178..347B}
\begin{equation}
    j_\mathrm{ISCO} = \frac{G M_\mathrm{BH}}{c} \frac{2}{3^{3/2}} \left[ 1 + 2 \left( \frac{3 c^2 r_\mathrm{ISCO}}{G M_\mathrm{BH}} -2 \right)^{1/2} \right] \, ,
\end{equation}
where $r_\mathrm{ISCO}$ is the radius of the ISCO for prograde equatorial orbits,
\begin{equation}
    r_\mathrm{ISCO} = \frac{G M_\mathrm{BH}}{c^2} \left\{ 3 + z_2 - \left[(3-z_1)(3+z_1+2z_2)\right]^{1/2} \right\} \, ,
\end{equation}
with $z_1 = 1 + (1-a^2)^{1/3}[(1-a)^{1/3}+(1+a)^{1/3}]$ and $z_2 = (3 a^2+z_1^2)^{1/2}$. From the disk formation condition $j(r,\theta) > j_\mathrm{ISCO}$ we can define the polar angle at which disk formation occurs as
\begin{equation}
    \theta_\mathrm{disk} = \arcsin{\left( \frac{j_\mathrm{ISCO}}{\Omega_\mathrm{shell}(r)r^2} \right)^{1/2}} \, . 
\end{equation}
The portion of the shell with $\theta < \theta_\mathrm{disk}$ will  collapse directly onto the BH on a dynamical timescale, $t_\mathrm{dyn}\simeq[r^3/(GM(r))]^{1/2}$, transferring $j(r,\theta)$ to the hole, while the portion of the shell with $\theta \geq \theta_\mathrm{disk}$ will form a disk and transfer only $j_\mathrm{ISCO}$ to the BH. The disk will be accreted on a viscous timescale $t_\nu \simeq \alpha_\nu^{-1} (R_\mathrm{circ}/H)^2 t_\mathrm{circ}$ assumed to be much smaller than $t_\mathrm{dyn}$ \citep{2019arXiv190404835B}. Here $H$ is the disk's scale height, $\alpha_\nu$ is the viscosity parameter and $t_\mathrm{circ}$ is the Keplerian orbital period at the accretion radius also known as circularisation radius $R_\mathrm{circ} = j_\mathrm{shell}^2 / (G M_\mathrm{BH})$. The collapsing shell contributes therefore to the AM of the BH by
\begin{equation}
\begin{split}
    J_\mathrm{shell} \equiv J_\mathrm{direct} + J_\mathrm{disk}
    = & \int_0^{\theta_\mathrm{disk}} m_\mathrm{shell} \Omega_\mathrm{shell}(r) \, r^2 \sin^3(\theta) \, \mathrm{d}\theta \, + \\
    & + \int_{\theta_\mathrm{disk}}^{\pi/2} m_\mathrm{shell} j_\mathrm{ISCO} \sin (\theta) \, \mathrm{d} \theta \, .
\end{split}
\end{equation}
The accretion disk has mass $m_\mathrm{disk}=m_\mathrm{shell} \cos (\theta_\mathrm{disk})$ and the mass-energy accreted onto the BH from the disk is $\Delta M_\mathrm{disk} = m_\mathrm{disk} [1 - 2G M_\mathrm{BH} /(3 c^2 r_\mathrm{ISCO})]^{1/2}$ \citep{1970Natur.226...64B,1974ApJ...191..507T}.

\begin{figure*}
\centering
\includegraphics[width=0.88\textwidth]{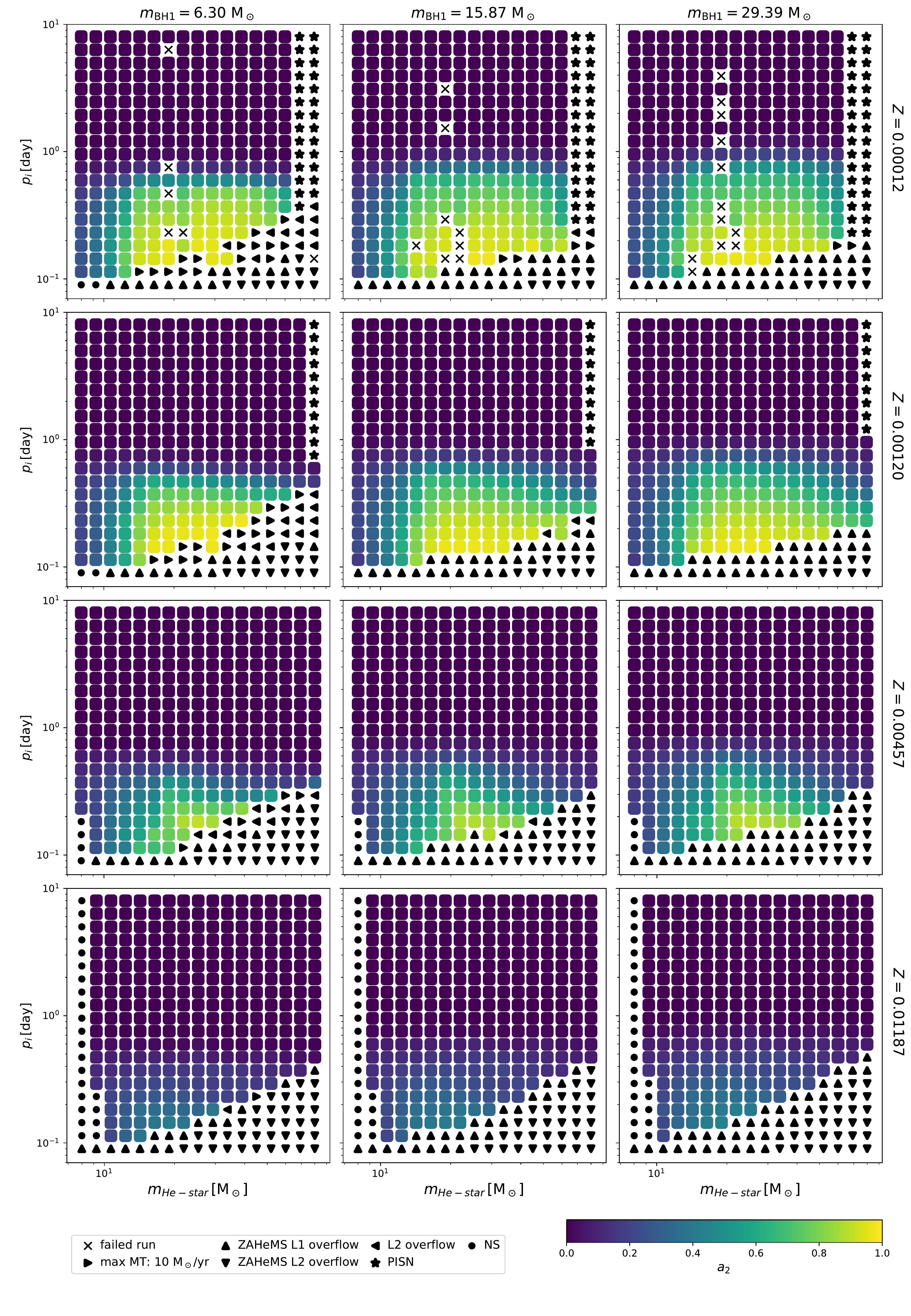}
 \caption{These are examples of two-dimensional slices of the four-dimensional grid for 4 different metallicities, $Z\in[0.00012,0.00120,0.00457,0.01187]$, and 3 different BH masses $m_\mathrm{BH}\in[6.3,15.87,29.38]~\mathrm{M}_\odot$. The final second-born BH spin value $a_2$ is coloured for each successful track according to the legend. Each successful run stopped because of carbon depletion (square markers) while other termination flags are shown in the bottom legend.
 }
 \label{Fig:grids-bgi}
\end{figure*}

\section{Grids of detailed BH--He-star models}\label{app:MESA-grids}
   
We use our detailed binary stellar models to cover the 4D parameter space defined by initial metallicity Z, black hole mass $m_\mathrm{BH1}$, He-star mass $m_\mathrm{He-star}$ and orbital period $p$. We run grids for 30 different metallicities ranging from $\log_{10}(Z) = -4.0$ to $\log_{10}(Z) = \log_{10}(1.5 Z_\odot) \simeq -1.593$ in steps of $\log_{10}(Z) \simeq 0.083$. For each metallicity, we run $11$ BH masses in the log-range $[2.5, 54.4] \, \mathrm{M}_\odot$ in steps of $\log_{10}(m_\mathrm{BH1}/\mathrm{M}_\odot) \simeq 0.134$. The lower limit is the smallest theoretical BH mass (in our model) while the maximum BH mass is chosen to have the second to last BH mass at $40~\mathrm{M}_\odot$ near the PISN cut. For each metallicity and BH mass we run 17 He-star masses in the log-range $[8, 80] \, \mathrm{M}_\odot$ in steps of $\log_{10}(m_\mathrm{He-star}/\mathrm{M}_\odot) \simeq 0.063$ and $20$ binary periods in the log-range $[0.09, 8] \, \mathrm{days}$ in steps of $\log_{10}(p/\mathrm{days}) \simeq 0.103$. We verified that smaller He-star masses do not lead to BH formation for any metallicity we consider. The maximum He-star mass and smallest orbital period where chosen to include any BH--He-star system produced by our \cosmic{} models. The maximal orbital period range ensure that we cover the parameter space well past the point where the BBH systems are merging within the Hubble time. 

In total, we calculated roughly $110,000$ new binary evolution sequences, as compared to about $16,000$ used in \citet{2020A&A...635A..97B}. The fraction of failed \mesa{} runs vary from $3\%$ to $0.1\%$ depending on metallicity. To minimise the loss of information created by the failed runs we rerun those models with an He-star mass increased by $5\%$ which reduced the failed runs by a factor of $3$. In Fig.~\ref{Fig:grids-bgi} we show the spin of the second-born BH for different two-dimensional slices of the four-dimensional parameter space for four different metallicities, $Z\in[0.00012,0.00120,0.00457,0.01187]$, and three different BH masses, $m_\mathrm{BH}\in[6.3,15.87,29.38]~\mathrm{M}_\odot$. We can see how the tides are more efficient at lower metallicities. This is because the stellar winds of He-stars are metallicity dependent $\propto (Z/Z_\odot)^{0.85}$ \citep{2001A&A...369..574V} and widen more efficiently the binaries at larger metallicities and hence reduce the impact of tides.

\begin{figure}
\centering
\includegraphics[width=\columnwidth]{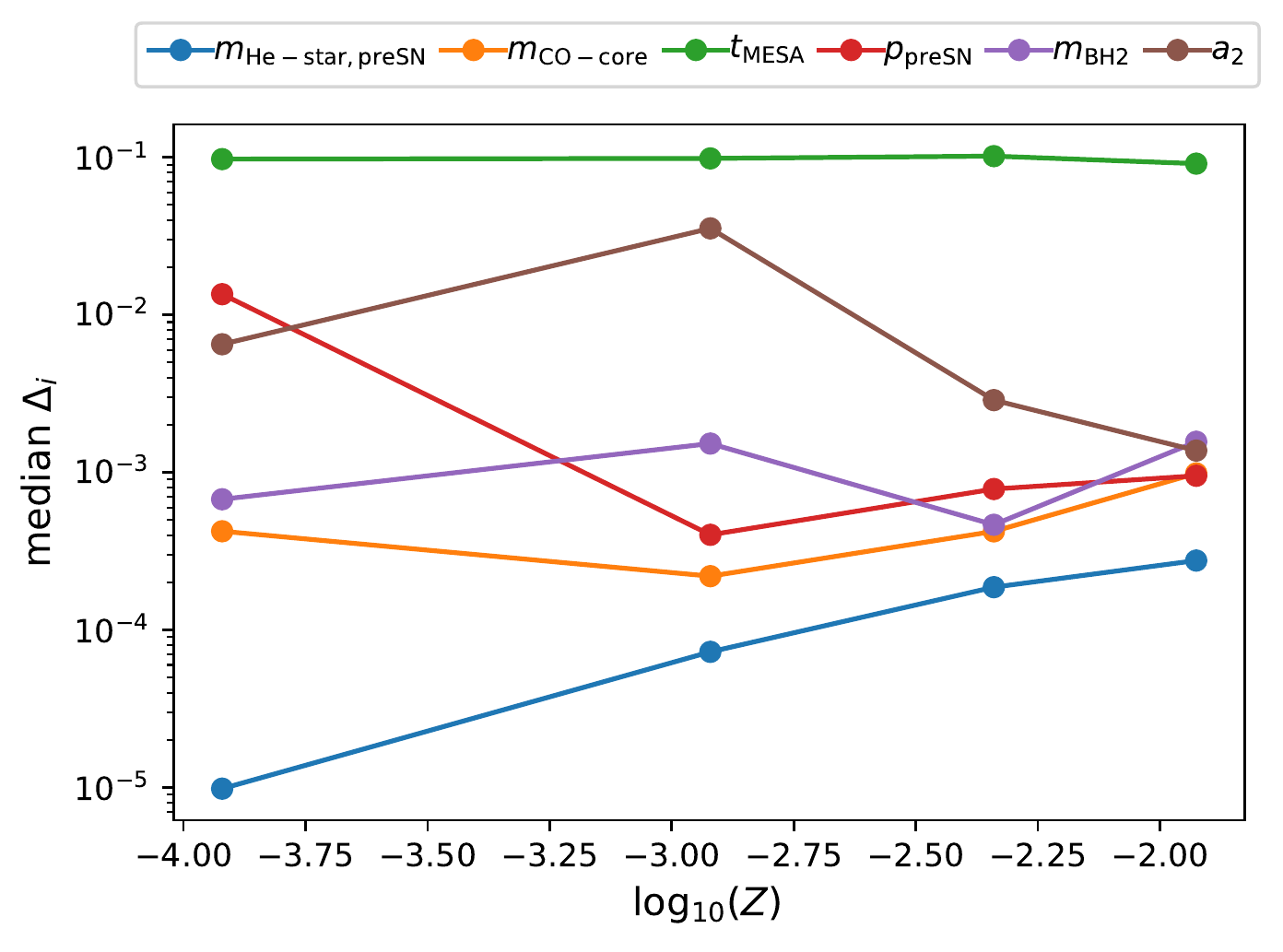}
 \caption{Median relative error as a function of metallicity of the log-transformed and re-scaled six quantities $B_i$: the He-star mass (blue) and its carbon-oxygen core mass (orange) before the supernova, the orbital period before the supernova (red), the second-born BH mass (violet), the spin of the second-born black hole (brown), and the lifetime of the BH–He-star binary (green).
 }
 \label{Fig:grids-errors}
\end{figure}
   
These grids were used to determine the final outcomes and final parameters of the late-end evolution stage of the binary systems through linear interpolation. Each metallicity is interpolated separately. We want to interpolate six quantities $B_i$: the He-star mass and its carbon-oxygen core mass before the supernova, the resulting BH mass, the orbital period before the supernova, the lifetime of the BH–He-star binary and the spin of the second-born BH. Before interpolating each quantity, we log-transformed it and re-scale it to the interval $[-1,1]$ to reduce the interpolation error. The interpolation itself relies
on building a Delaunay triangulation of the input data points followed by barycentric linear interpolation over the vertices of the (hyper)triangle containing the location of interest. To test the accuracy of the interpolation: we computed around $1000$ new \mesa{} tracks for each of the four metallicities shown in Fig.~\ref{Fig:grids-bgi} and calculate the relative error of each transformed and rescaled quantity $X_i\equiv \log_{10}(B_i)^{[-1,1]}$ as $\Delta_i = |X_\mathrm{true,i}-X_\mathrm{interp,i}|/X_\mathrm{true,i}$. In Fig.~\ref{Fig:grids-errors} we show the median relative errors of these quantities as a function of metallicity. In the median calculation we exclude all the systems not becoming BBHs. Half of the quantities have median relative errors independent of metallicity with the exception of the second-born BH spin, He-star mass and orbital period before supernova. The decrease of median error for $a_2$ and $p_\mathrm{preSN}$ is explained by the fact that at high metallicities the orbits widen more, neutralising tides and resulting in systems with zero spin. On the other hand, the median error of $m_\mathrm{He-star}$ increases because at higher metallicities this quantity does not depend linearly on the initial He star mass which is caused by stellar winds that cause the He-star to lose a non-negligible amount of mass. The largest relative median error is the lifetime of the BH--He-star system. This is not a problem by itself as the delay time between the binary formation and merger is dominated by the GW inspiral which is many order of magnitudes larger than this timescale. The increase of interpolation accuracy compared to \cite{2020A&A...635A..97B} is due to the larger data set used here, the fact that we use a regular grid and interpolate each metallicity independently.

\section{Model comparison}\label{app:model-comparison}

We use Bayesian hierarchical modelling to determine the likelihood of observing $N$ independent GW events $\{x_i\}_{i=1}^N$ given an astrophysical model described by a set of parameters $\vec{\lambda}$ \citep[e.g.][]{2019MNRAS.486.1086M}. We assume that each GW event is an independent observation and is characterised by a set of physical parameters $\vec{\theta}$,
\begin{equation}
\begin{split}
    p(\{x_i\}_{i=1}^N \, | \, \vec{\lambda}) &= \prod_{i=1}^N \frac{p(x_i \, | \, \vec{\lambda})}{\xi(\vec{\lambda})} = \prod_{i=1}^N \frac{1}{\xi(\vec{\lambda})} \int p(x_i \, | \, \vec{\theta}) \, p(\vec{\theta} \, | \, \vec{\lambda}) \, \,\mathrm{d}\vec{\theta} = \\
    &=  \prod_{i=1}^N  \frac{p(x_i)}{\xi(\vec{\lambda})} \int \frac{p(\vec{\theta} \, | \, x_i)}{p(\vec{\theta})} p(\vec{\theta} \,|\, \vec{\lambda}) \, \mathrm{d}\vec{\theta} \, ,
\end{split}
\label{eq:xi_sample}
\end{equation}
where we have marginalised over the physical parameters of the individual events and used Bayes' theorem to obtain the final line. Here $p(\vec{\theta})$ is the prior on the physical parameters that are used to generate the posterior samples. The normalisation factor $\xi(\vec{\lambda}) = \int p_\mathrm{det}(\vec{\theta})p(\vec{\theta}|\vec{\lambda}) \mathrm{d}\vec{\theta}$ accounts for the overall probability of making an observation given a particular choice $\vec{\lambda}$. Since in GWTC-2 we have samples drawn from the posterior $p(\vec{\theta} \, | \, x_i)$, we can approximate posterior-weighted integrals as a sum over samples as
\begin{equation}
    p(\{x_i\}_{i=1}^N \, | \, \vec{\lambda}) \simeq \prod_{i=1}^N \frac{p(x_i)}{\xi(\vec{\lambda})} \, \frac{1}{S} \sum_{k=1}^S \frac{p(\vec{\theta}_k \, | \, \vec{\lambda})}{p(\vec{\theta}_k)} \, ,
    \label{eq:likelihood}
\end{equation}
where $S$ the number of posterior samples. Again, $\vec{\theta}_k$ are the astrophysical parameters drawn from the GW posterior distribution and $p(\vec{\theta}_k \, | \, \vec{\lambda})$ are their likelihood given by our astrophysical model (underlying distribution). Similarly, the normalisation constant $\xi(\vec{\lambda})$ can be approximated as a sum over the weighed underlying BBH distribution $p(\vec{\theta}|\vec{\lambda})$ as
\begin{equation}
    \xi(\vec{\lambda}) \simeq \frac{1}{T \sum_{j=1}^T w_j} \sum_{j=1}^T p_\mathrm{det}(\vec{\theta}_j) \, w_j \, ,
\end{equation}
where $\vec{\theta}_j$ is the set of parameters describing the BBH $j$ with a cosmological weight $w_j$ given by the argument of the summation in Eq. (\ref{eq:Rdet}) and $T$ the total number of samples.
Finally, to approximate the probability density functions of each event $p(\vec{\theta})$ and each model $p(\vec{\theta} | \vec{\lambda})$ we use a three-dimensional kernel density estimator (KDE) where we graphically verify the accuracy by comparing random draws from the KDE with the real marginalised one-dimensional and two-dimensional sample distributions.

We can now compare two models, $M_1$ and $M_2$, described by $\vec{\lambda}_1$ and $\vec{\lambda}_2$, respectively. The amount by which the data supports a specific model is described by the BF, defined as:
\begin{equation}
    \mathrm{BF}_{12} \equiv \frac{p(\{x_i\}_{i=1}^N \, | \, \vec{\lambda}_1)}{p(\{x_i\}_{i=1}^N \, | \, \vec{\lambda}_2) } \, .
    \label{eq:BF}
\end{equation}
In the ratio of the two likelihoods the multiplicative constant $p(x_i)$ is canceled out, leaving us with all the information required to compute the BF. This factor indicates whether any model is favoured or disfavoured by the data compared to another. Values larger than 1 favour the model $M_1$ while values smaller than 1 favours the model $M_2$. In our analysis, we adopt the convention of having $M_2$ as our reference model.

\section{Model results}\label{app:model-results}

\begin{figure*}
\vspace{-0.5cm}
\thisfloatpagestyle{empty}
    \centering
    \includegraphics[width=0.4\textwidth]{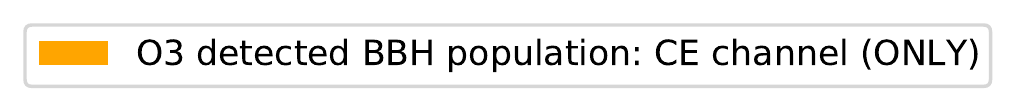} \\
    \includegraphics[width=0.75\textwidth]{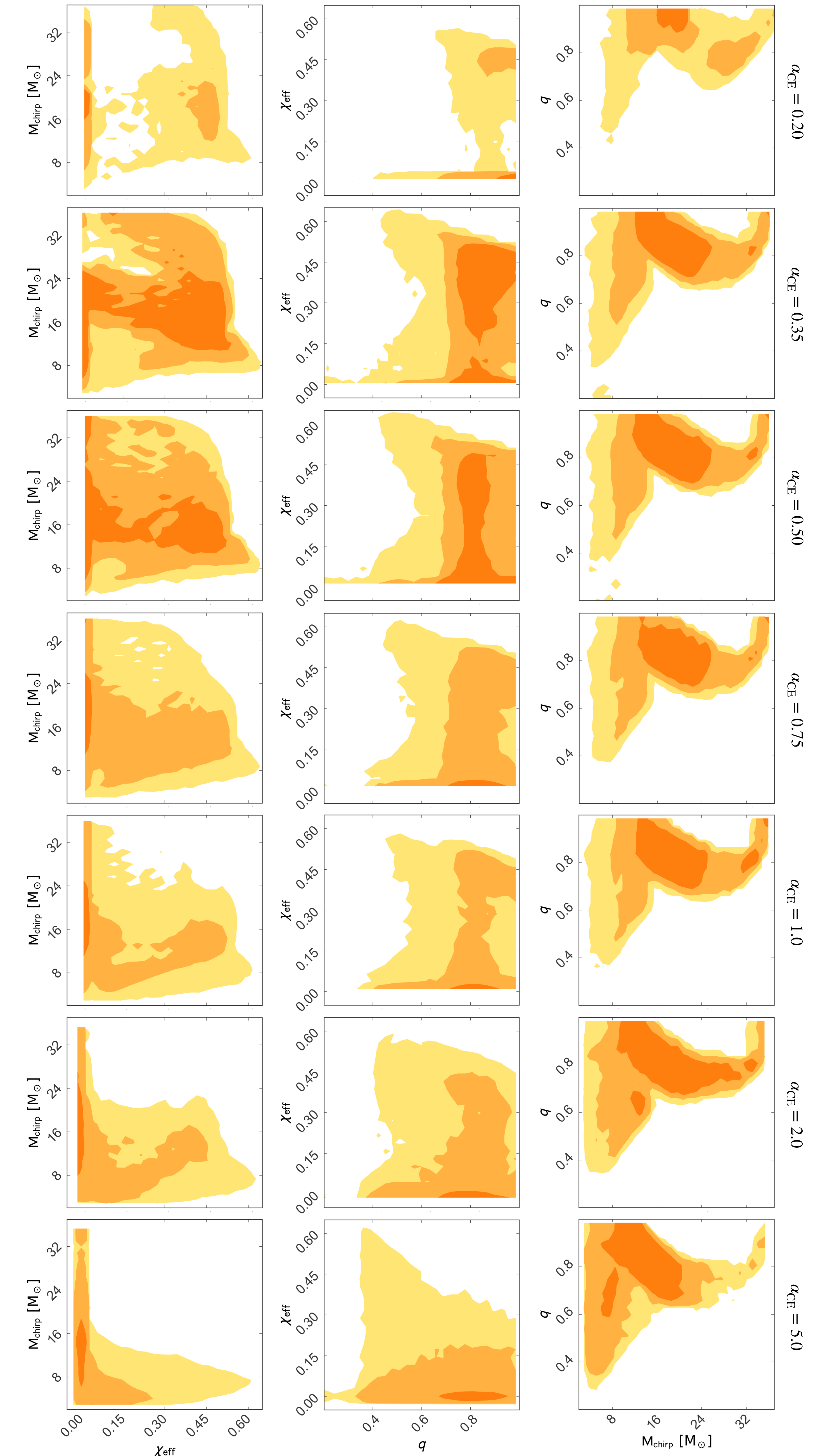}
    \caption{Model predictions for the O3 detected BBH population of the CE channel for different $\alpha_\mathrm{CE}$ values according to the legend, in orange. We show the joint distributions of chirp mass $M_\mathrm{chirp}$, effective inspiral spin parameter $\chi_\mathrm{eff}$ and binary mass ratio $q$. Lighter colours represent larger contour levels of $68\%$, $95\%$ and $99\%$, respectively, constructed with \texttt{pygtc} \citep{Bocquet2016}. All histograms are plotted with 30 bins without any bin smoothing.
    }
     \label{fig:models-O3-CE-alphas}
\end{figure*}

\begin{figure*}
\vspace{-0.5cm}
\thisfloatpagestyle{empty}
    \centering
    \includegraphics[width=0.4\textwidth]{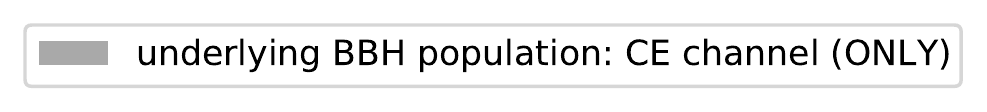} \\
    \includegraphics[width=0.75\textwidth]{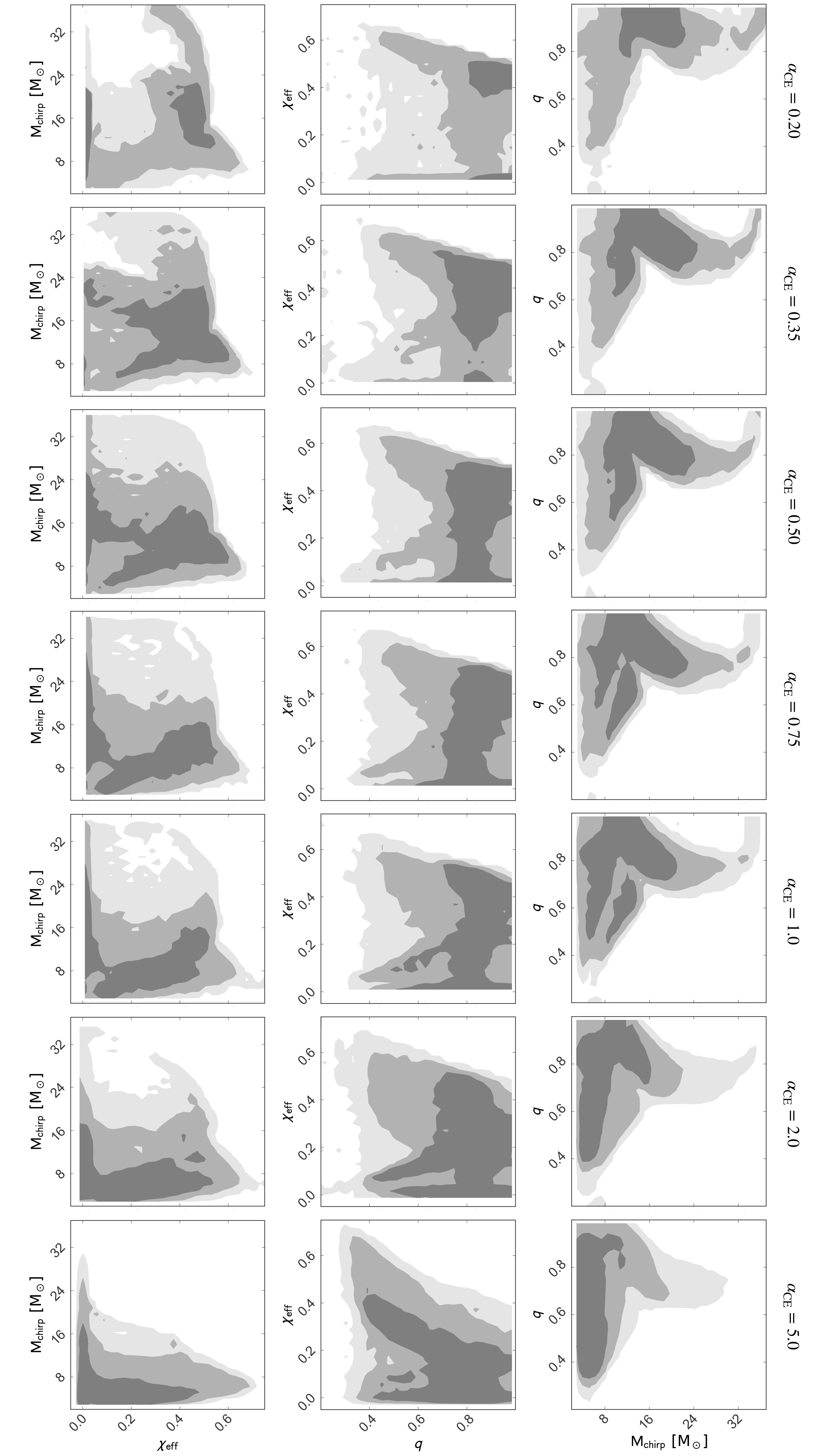}
    \caption{Model predictions for the underlying (intrinsic) BBH population of the CE channel for different $\alpha_\mathrm{CE}$ values according to the legend, in grey. We show the joint distributions of chirp mass $M_\mathrm{chirp}$, effective inspiral spin parameter $\chi_\mathrm{eff}$ and binary mass ratio $q$. Lighter colours represent larger contour levels of $68\%$, $95\%$ and $99\%$, respectively, constructed with \texttt{pygtc} \citep{Bocquet2016}. All histograms are plotted with 30 bins without any bin smoothing.
    }
     \label{fig:models-inf-CE-alphas}
\end{figure*}

\begin{figure*}
    \centering
    \includegraphics[width=0.4\textwidth]{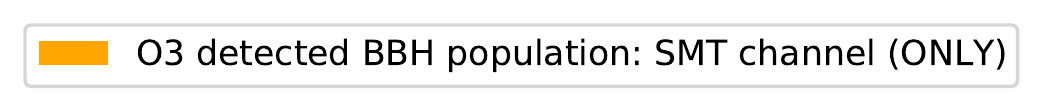} \\
    \includegraphics[width=0.75\textwidth]{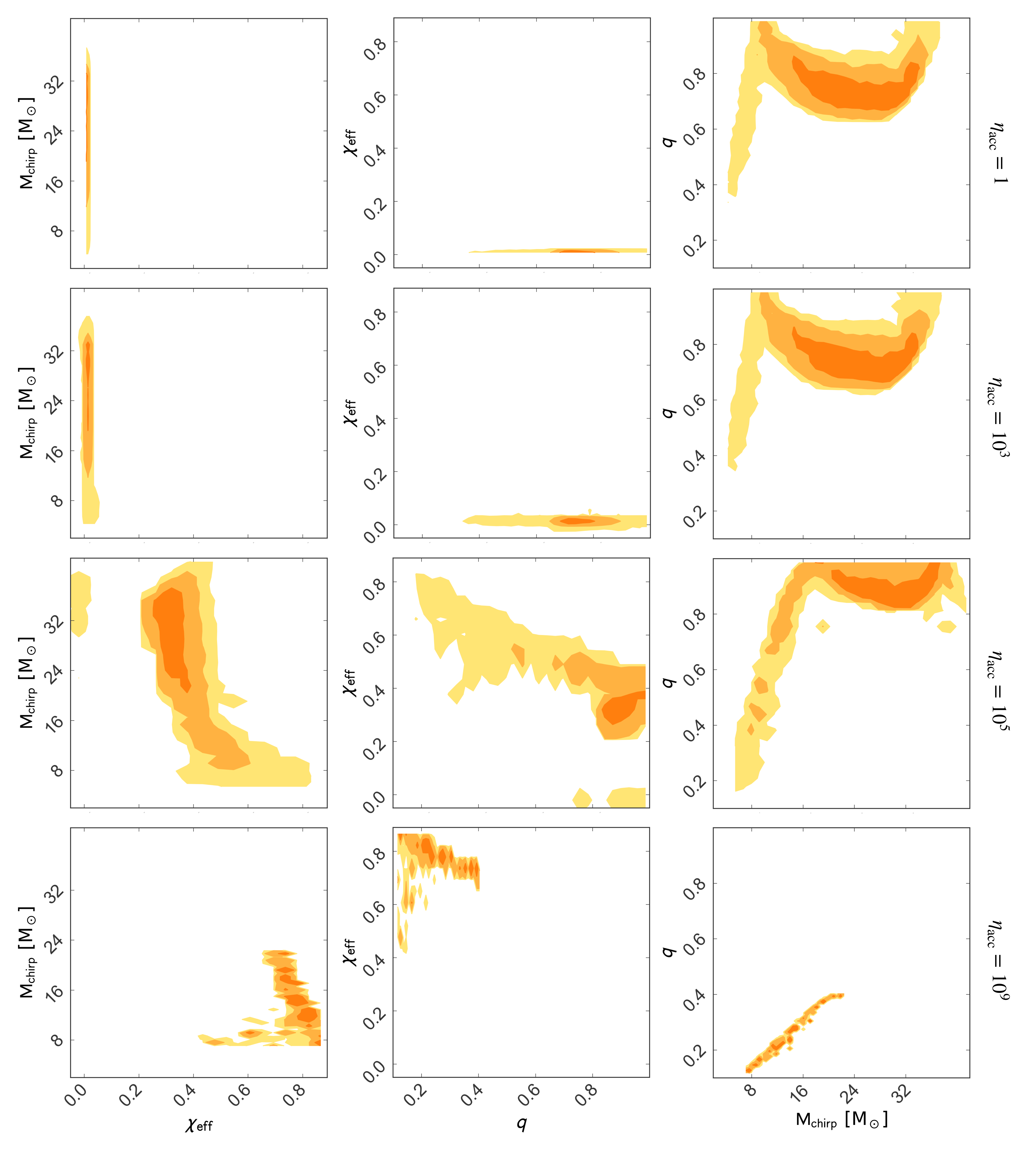}
    \caption{Model predictions for the O3 detected BBH population of the SMT channel for different $\eta_\mathrm{acc}$ values according to the legend, in orange. We show the joint distributions of chirp mass $M_\mathrm{chirp}$, effective inspiral spin parameter $\chi_\mathrm{eff}$ and binary mass ratio $q$. Lighter colours represent larger contour levels of $68\%$, $95\%$ and $99\%$, respectively, constructed with \texttt{pygtc} \citep{Bocquet2016}. All histograms are plotted with 30 bins without any bin smoothing.
    }
    \label{fig:models-O3-SMT-etas}
\end{figure*}

\begin{figure*}
    \centering
    \includegraphics[width=0.4\textwidth]{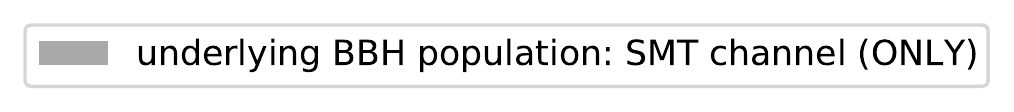} \\
    \includegraphics[width=0.75\textwidth]{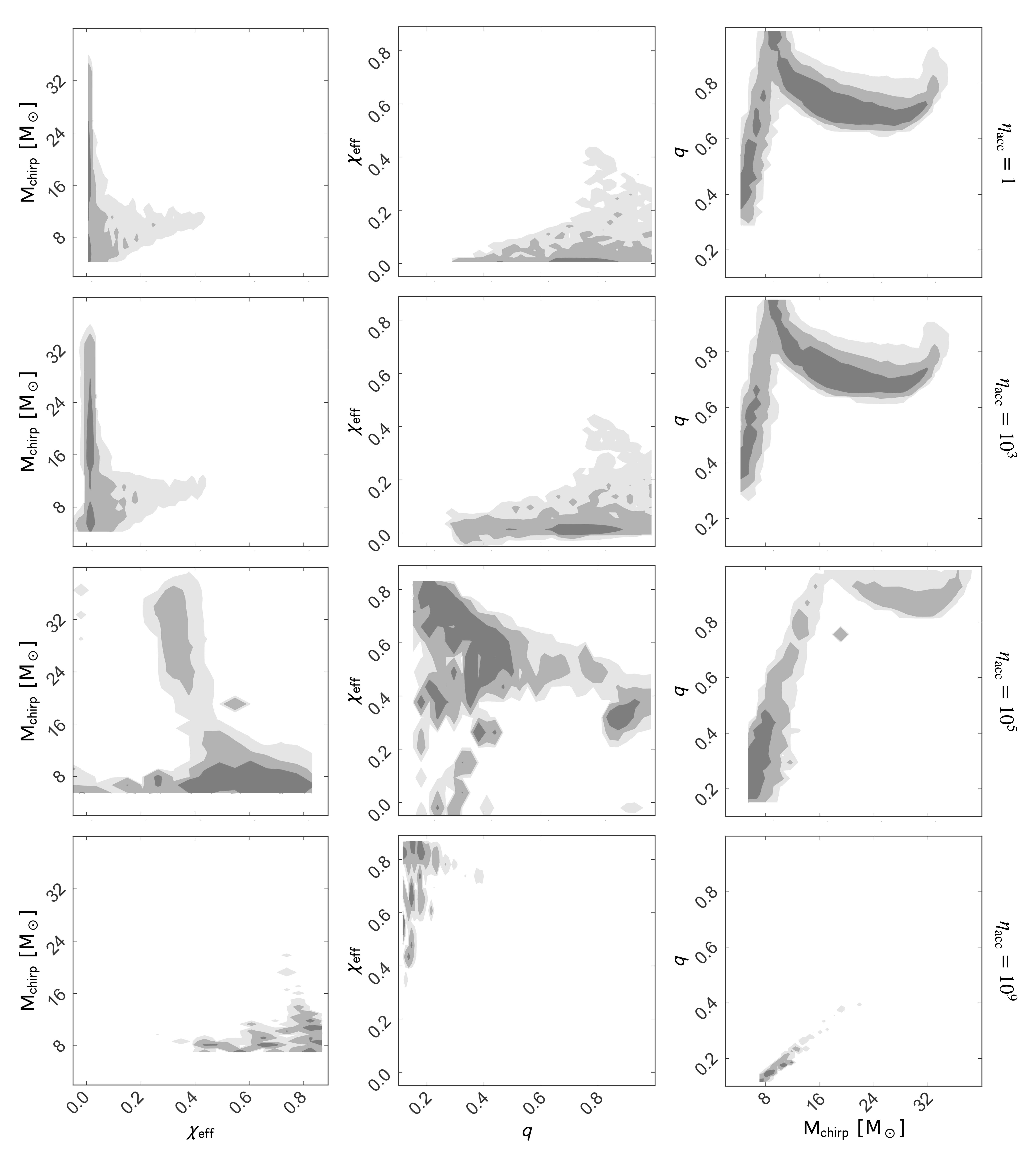}
    \caption{Model predictions for the underlying (intrinsic) BBH population of the SMT channel for different $\eta_\mathrm{acc}$ values according to the legend, in grey. We show the joint distributions of chirp mass $M_\mathrm{chirp}$, effective inspiral spin parameter $\chi_\mathrm{eff}$ and binary mass ratio $q$. Lighter colours represent larger contour levels of $68\%$, $95\%$ and $99\%$, respectively, constructed with \texttt{pygtc} \citep{Bocquet2016}. All histograms are plotted with 30 bins without any bin smoothing.
    }
    \label{fig:models-inf-SMT-etas}
\end{figure*}

Here we present some extra figures which were not included in the paper because of their size. Fig.~\ref{fig:models-O3-CE-alphas} and ~\ref{fig:models-inf-CE-alphas} show the combined distributions of the main GW observables \chieff, \mchirp and $q$ of the CE channel for the detected and underlying BBH population, respectively, for different $\alpha_\mathrm{CE}$ values. Similarly, Fig.~\ref{fig:models-O3-SMT-etas} and Fig.~\ref{fig:models-inf-SMT-etas} show the combined distributions of these observables of the SMT channel for the detected and underlying BBH population, respectively, for different $\eta_\mathrm{acc}$ values.

\end{document}